\documentclass[prb,reprint, amsmath, amsymb, twocolumn,superscriptaddress,floatfix,preprintnumbers]{revtex4-2}

\usepackage{amsmath,amssymb,graphicx}
\usepackage{ mathrsfs }
\usepackage{algorithm}
\usepackage{algpseudocode}
\usepackage{xr-hyper}
\usepackage[colorlinks=true, urlcolor=blue,citecolor=blue,anchorcolor=blue]{hyperref}
\usepackage{xcolor}
\usepackage{ulem}

\newcommand{\ket}[1]{\left|#1\right\rangle}
\newcommand{\bra}[1]{\left\langle#1\right|}

\newcommand{\der}[2]{\frac{\mathrm{d} #1}{\mathrm{d} #2}}

\newcommand{\pder}[2]{\frac{\partial #1}{\partial #2}}

\begin{document}

\title{Lefschetz thimble quantum Monte Carlo for spin systems}

\author{T.~C.~Mooney}
\affiliation{Joint Center for Quantum Information and Computer Science, NIST/University of Maryland, College Park, Maryland 20742, USA}
\affiliation{Joint Quantum Institute, NIST/University of Maryland, College Park, Maryland 20742, USA}

\author{Jacob~Bringewatt}
\affiliation{Joint Center for Quantum Information and Computer Science, NIST/University of Maryland, College Park, Maryland 20742, USA}
\affiliation{Joint Quantum Institute, NIST/University of Maryland, College Park, Maryland 20742, USA}

\author{Neill~C.~Warrington}
\affiliation{Institute for Nuclear Theory, University of Washington, Seattle, Washington 98195, USA}

\author{Lucas~T.~Brady}
\affiliation{Quantum Artificial Intelligence Laboratory, NASA Ames Research Center, Moffett Field, California 94035, USA}
\affiliation{KBR, 601 Jefferson St., Houston, TX 77002, USA}

\preprint{INT-PUB-22-015}

\date{\today}

\begin{abstract}
Monte Carlo simulations are useful tools for modeling quantum systems, but in some cases they suffer from a sign problem, leading to an exponential slow down in their convergence to a value. While solving the sign problem is generically NP-hard, many techniques exist for mitigating the sign problem in specific cases; in particular, the technique of deforming the Monte Carlo simulation's plane of integration onto Lefschetz thimbles (complex hypersurfaces of stationary phase) has seen significant success in the context of quantum field theories. We extend this methodology to spin systems by utilizing spin coherent state path integrals to re-express the spin system's partition function in terms of continuous variables. Using some toy systems, we demonstrate its effectiveness at lessening the sign problem in this setting, despite the fact that the initial mapping to spin coherent states introduces its own sign problem. The standard formulation of the spin coherent path integral is known to make use of uncontrolled approximations; despite this, for large spins they are typically considered to yield accurate results, so it is somewhat surprising that our results show significant systematic errors. Therefore, possibly of independent interest, our use of Lefschetz thimbles to overcome the intrinsic sign problem in spin coherent state path integral Monte Carlo enables a novel numerical demonstration of a breakdown in the spin coherent path integral. 
\end{abstract}
\maketitle

\section{Introduction}
The sign problem in quantum Monte Carlo (QMC) is a major impediment to efficiently simulating quantum systems classically. While the sign problem is basis dependent, it is NP-hard, in general, to find a basis with no sign problem \cite{troyer_computational_2005, marvian_computational_2019}. Still, there are numerous heuristic strategies for solving the sign problem in specific cases \cite{hen_resolution_2019, Li_2015}, or mitigating it in others, reducing the severity of the exponential slow down without eliminating it \cite{wan2020mitigating,hangleiter_easing_2020}.

The sign problem arises via the protocol of reweighting, in which one pushes the quantum quasiprobability distribution's phase onto the observable in order to get proper probabilities with which to conduct Monte Carlo simulations. Specifically, for some (complex) quasiprobability distribution $\rho = p e^{i \theta }$, we have that the expectation value of an observable $O$ is given by:
\begin{align}
    \langle O \rangle_{\rho} &= \frac{\langle O e^{i \theta}\rangle_p }{\langle e^{i\theta}\rangle _p}.
\end{align}
This reweighting allows non-positive and non-real quasiprobabilities to be sampled via Monte Carlo methods, but the method falls apart when the phase is highly oscillatory. This is the case because for a Monte Carlo sampling of an $n-$particle system repeated $m$ times, the relative error in the phase can be written as 
\begin{align}
    \frac{\Delta e^{i\theta}}{\langle e^{i\theta}\rangle_p} &= \frac{e^{\mathcal{O}(\beta n)}}{\sqrt{m}},
\end{align}
where $\Delta e^{i\theta}$ is the standard deviation of $e^{i\theta}$ and $\beta$ is the inverse temperature \cite{troyer_computational_2005}. Thus, to keep a constant level of relative error in the Monte Carlo simulation, $m$ must scale exponentially with both particle number and inverse temperature, neither of which is desirable.

The sign problem can be avoided by Hamiltonians that are stoquastic (where all the off-diagonal entries are real and non-positive) \cite{bravyi2006complexity} or, more generally, by Hamiltonians that are in Vanishing Geometric Phase form \cite{jarret2018hamiltonian,Hen_2021}. However, a variety of interesting quantum many-body systems are not of this form, motivating efforts to mitigate the sign problem. For instance, the spin-1/2 Heisenberg model on a Kagome lattice is a canonical example.

In addition to the clear relevance for condensed matter physics, attempts to mitigate the sign problem are of interest in regards to the study of analog quantum computation. In a quantum computing context, stoquastic Hamiltonians play into quantum complexity theory \cite{bravyi2008complexity,bravyi2006MerlinArthurGA}. In adiabatic quantum computing specifically, there is a large body of evidence that local sign-problem free Hamiltonians do not possess quantum advantage over classical computing \cite{crosson_2021,Bringewatt_2020} and that quantum advantage with stoquastic Hamiltonians requires contrived non-local systems \cite{hastings2020power}.
Adiabatic quantum computation with general Hamiltonians is known to be universal \cite{aharonov2008adiabatic}, and therefore, it is expected that the quantum advantage over classical computation arises for Hamiltonians that exhibit a sign problem. Nonetheless, many classical techniques are known to address the sign problem in certain instances, motivating us to consider for which Hamiltonians the sign problem makes classical simulation truly intractable.

Lefschetz thimble methods, based on Picard-Lefschetz theory \cite{pham1983vanishing}, are one promising strategy for mitigating the sign problem \cite{witten2010new, witten2011analytic}. These methods are a higher dimensional analogue of the stationary phase method, deforming an integral into complex space so that it sits on a manifold of stationary phase. This mostly eliminates the rapid phase oscillations which lead to the sign problem.

This approach has primarily been developed in the context of quantum field theories, both bosonic \cite{Cristoforetti_2012, cristoforetti2013montecarlo, mukherjee2013metropolis, Fujii_2013, aarts2013lefschetz, Cristoforetti_2014} and fermionic \cite{aarts2014some,kanazawa2015structure, fujii2015lefschetz, alexandru2016mc, alexandru2017mc, alexandru2018fermions, direnzo2018one, alexandru2018finite}. Most such field theoretic problems are immediately amenable to the Lefschetz thimble approach, as the relevant variables are continuous and, thus, easily complexified. The approach has also been applied to the Hubbard model \cite{mukherjee2014lefschetz, tanizaki2016lefshetz, ulybyshev2020lefschetz}, where prior to applying the techniques one must map the discrete partition function to a functional integral over continuous variables via a Hubbard-Stratonovich transformation \cite{stratonovich1957method, hubbard1959calculation}. Similarly, for spin systems, one must also map the problem to continuous variables. This was recently done by mapping spins onto complex fermions, but this approach was limited to 2-body interactions between spin-1/2 particles~\cite{mishchenko_quantum_2021}. 

In this paper, we adapt Lefschetz thimble Monte Carlo methods to generic spin systems by utilizing spin coherent states to map the partition function for a spin system into a continuous variable setting. Using these continuous variable models for spin, we implement a Lefschetz thimble method to mitigate the sign problem in these spin systems. This provides a potential path toward simulating a larger array of quantum systems using classical methods. 

While our application of the Lefschetz thimble techniques is effective at mitigating the sign problem, unfortunately, our results show significant systematic errors compared to those obtained by exact diagonalization. This error is due to uncontrolled approximations made in the standard formulation of the spin coherent path integral~\cite{shibata_note_1999, auerbach}. While it has previously been demonstrated analytically that these approximations can lead to inaccurate results~\cite{galitski2011breakdown}, our use of Lefschetz thimbles provides, to our knowledge, the first numerical demonstration of such a breakdown. This is because standard spin coherent path integral Monte Carlo introduces its own sign problem, making such calculations extremely expensive without using our techniques to mitigate the sign problem. 

Our novel numerical demonstration of such a breakdown is an interesting result in its own right, but for the purposes of applying Lefschetz thimbles to study large spin systems with a sign problem, it indicates a hurdle to be overcome. The success of these techniques at mitigating the sign problem demonstrate that better understanding and controlling the systematic errors that arise in spin coherent path integrals~\cite{kordas2016coherent, ranccon2020hubbard}, is a worthwhile goal as it would immediately unlock a powerful new technique for studying spin systems with a sign problem.

\section{Lefschetz Thimbles}
Consider an integral of the form 
\begin{equation}
    \mathcal{Z}=\int_{\mathbb{R}^n}\mathrm{d}^n\mathbf{x}\,e^{-\mathcal{S}(\mathbf{x})},
\end{equation}
where, here, $n$ is the number of degrees of freedom, $\mathbf{x}\in\mathbb{R}^n$ (boldface denotes a vector) are the state variables, and $\mathcal{S}$ is the action. For a complex action $\mathcal{S}$, this integrand can be highly oscillatory, and, therefore, can suffer from the sign problem when numerically integrated via QMC. The core of the Lefschetz thimble approach to mitigating this sign problem is to promote $\mathcal{S}(\mathbf{x}):\mathbb{R}^n\to \mathbb{C},$ to a holomorphic function $\mathcal{S}(\mathbf{z}): \mathbb{C}^n\to\mathbb{C},$ and deform the original integration region, called an $n$-\textit{cycle}, to a collection of cycles of stationary phase for the action, on which the imaginary part of the action is constant. On such stationary phase cycles, called \textit{Lefschetz thimbles}, there is no sign problem.

To formally define the stationary phase cycles and to systematize deformation into them, we introduce the concept of a \textit{holomorphic flow.} Let $\tau$ be the flow-time parameter, a non-physical parameterization of our deformation in the complex hyperplane. We write the result of flowing an initial integration point, $\mathbf{z}_0$, for flow-time $\tau$ as $\varphi(\mathbf{z}_0;\tau): \mathbb{C}^n\times \mathbb{R}\to \mathbb{C}^n.$ Letting $\mathbf{z}(\tau)\equiv \varphi(\mathbf{z}_0;\tau),$ we define the holomorphic flow via the differential equation
\begin{equation}\label{eq:flow}
    \der{z_i}{\tau} = \overline{\pder{\mathcal{S}}{z_i}},
\end{equation}
where the bar denotes complex conjugation.
There are a few important things to note about Eq.~(\ref{eq:flow}). First, the holomorphic flow is such that the real part of the action monotonically increases under the flow, whereas the imaginary part of the action is constant. This can be confirmed by observing that
\begin{align*}
    \der{\mathcal{S}}{\tau}&= \sum_i \pder{\mathcal{S}}{z_i}\der{z_i}{\tau}= \sum_i \pder{\mathcal{S}}{z_i}\overline{\pder{\mathcal{S}}{z_i}} \geq 0,
\end{align*}
and that $\mathrm{d} \mathcal{S}/\mathrm{d} \tau$ is real. Alternatively one could simply observe that the holomorphic flow is: (a) the gradient flow of $\mathrm{Re}\,\mathcal{S}$, and, therefore, it is the path of steepest ascent for $\mathrm{Re}\,\mathcal{S}$; and (b) the Hamiltonian flow for a ``Hamiltonian'' given by $\mathrm{Im}\,\mathcal{S}$, and, consequently, $\mathrm{Im}\,\mathcal{S}$ is conserved under the flow. Second, the critical points of the action (i.e., where $\partial\mathcal{S}/\partial z_i=0$ for all $i$) are stationary under the flow. Finally, the Jacobian $J$ corresponding to the change of variables $\mathbf{z}_0 \rightarrow \varphi(\mathbf{z}_0;\tau)$ satisfies its own flow equation,
\begin{equation} \label{eq:Jacobian_flow}
    \der{J}{\tau} = \overline{H J},
\end{equation}
where $H=\left[\frac{\partial^2\mathcal{S}}{\partial z_i\partial z_j}\right]$ is the Hessian of $\mathcal{S}.$

The flow equations allow us to formally define two interrelated stationary phase $n$-cycles: the \textit{Lefschetz thimble}, and the \textit{anti-thimble.} Let $\{p_\sigma\vert \sigma \in \Sigma\}$ be the collection of critical points of $\mathcal{S},$ with some indexing set $\Sigma.$ Then, we can define the \textit{Lefschetz thimble attached to $p_\sigma,$} denoted $\mathcal{J}_\sigma$, to be the collection of all points that flow away from $p_\sigma$ under Eq.~(\ref{eq:flow}). Mathematically:
\begin{equation}
    \mathcal{J}_\sigma = \{ \mathbf{z} \vert \varphi(\mathbf{z};-\infty) = p_\sigma\}.
\end{equation}
Similarly, the \textit{anti-thimble attached to $p_\sigma,$} written $\mathcal{K}_\sigma$, is defined as the set of all points that flow to $p_\sigma$. That is: 
\begin{equation}
    \mathcal{K}_\sigma = \{ \mathbf{z} \vert \varphi(\mathbf{z};\infty) = p_\sigma\}.
\end{equation}
Notice that $\mathcal{J}_\sigma$, in addition to being a stationary phase cycle of $\mathcal{S}$, is also the steepest ascent cycle of $\mathcal{S}$ from $p_\sigma$. This means that we know definitively that $\mathcal{J}_\sigma$ is a convergent integration cycle for $e^{-\mathcal{S}}$, as the boundaries of the integral go to zero as quickly as possible. On the other hand, $\mathcal{K}_\sigma$ is a cycle of steepest descent, and, therefore, divergent.

The anti-thimbles $\mathcal{K}_\sigma$ serve their purpose, however, by helping to identify the set of thimbles $\mathcal{J}_\sigma$ that correspond to our initial integration contour $\mathbb{R}^n$. In particular, if some modest conditions are met \cite{pham1983vanishing, witten2011analytic}, the integral of $e^{-\mathcal{S}}$ over any convergent integration cycle can be deformed into an integral over a linear combination of $\mathcal{J}_\sigma$'s. Given our initial integration cycle $\mathbb{R}^n$ we have that
\begin{equation}\label{eq:pairing}
    \mathbb{R}^n \simeq \sum_{\sigma'} \langle \mathcal{K}_{\sigma'},\mathbb{R}^n\rangle \mathcal{J}_{\sigma'}
\end{equation}
up to equality of integration, where $\langle A, B\rangle$ is the intersection pairing of $n$-cycles $A$ and $B,$ i.e. the number of isolated intersections between $A$ and $B$. 

Provided we can appropriately identify (or approximately identify) the correct intersection pairings, Eq.~(\ref{eq:pairing}) provides us a pathway to ameliorating the sign problem by integrating over the appropriate linear combination of Lefschetz thimbles instead of $\mathbb{R}^n$. While a variety of numerical treatments have shown this approach to be useful in many contexts \cite{mukherjee2013metropolis, renzo2015thimble, fujii2015lefschetz, alexandru2016mc, alexandru2016sign, alexandru2017mc, fukuma2017parallel, alexandru2017schwinger}, it is important to emphasize that this change of integration contour does not fully solve the sign problem. One limitation is that, during the deformation process, the phase of the integrand will pick up a contribution from the Jacobian (the so-called residual phase), which may introduce a sign problem of its own \cite{Cristoforetti_2012, alexandru_complex_2020}. Furthermore, if multiple thimbles contribute to the partition function, a sign problem can still arise due to the presence of relative phases between different thimbles; such relative phases arise because, while $\mathrm{Im}\,\mathcal{S}$ is constant on individual thimbles, it is not so between different thimbles \cite{alexandru_complex_2020}. Despite these caveats, in many cases of interest these issues have not been fatal and the Lefschetz thimble approach has seen great success \cite{Cristoforetti_2012, cristoforetti2013montecarlo, mukherjee2013metropolis, Fujii_2013, aarts2013lefschetz, Cristoforetti_2014, aarts2014some,kanazawa2015structure, fujii2015lefschetz, alexandru2016mc, alexandru2017mc, alexandru2018fermions, direnzo2018one, alexandru2018finite, mukherjee2014lefschetz, tanizaki2016lefshetz, ulybyshev2020lefschetz, mishchenko_quantum_2021, direnzo2018one}, motivating its further application to sign problems in spin systems.

\section{Algorithms}
Now that we have introduced an analytic framework for addressing the sign problem, we consider the use of Lefschetz thimbles in QMC algorithms. A number of such algorithms have been introduced \cite{mukherjee2013metropolis, renzo2015thimble, fujii2015lefschetz, alexandru2016mc, alexandru2016sign, alexandru2017mc, fukuma2017parallel, alexandru2017schwinger}; here we consider a particular approach called the \textit{generalized thimble method,} first proposed in Ref.~\cite{alexandru2016sign}. This algorithm is of particular interest since it does not require \textit{a priori} knowledge of the critical points of the action in the complexified space, and, in principle, guarantees sampling over all relevant thimbles. 

The key to this algorithm is the observation that 
\begin{equation} \varphi(\mathbb{R}^n; \infty) = \sum_{\sigma'} \langle \mathcal{K}_{\sigma'},\mathbb{R}^n\rangle \mathcal{J}_{\sigma'}, \end{equation}
which can be intuitively understood by noting that the only structures that the flow can ``get stuck'' on are the thimbles, and all points not flowing to the thimbles will flow out to infinity, where they, too, will eventually arrive at the thimbles. Thus, for some sufficiently large time $\tau,$ a flow of $\mathbb{R}^n$ will approach the appropriate ensemble of thimbles, ameliorating the sign problem. 

\begin{figure}
\begin{algorithm}[H]
\caption{Generalized thimble method}\label{alg:LTMC}
\begin{algorithmic}[1]
\Require $N,\tau\geq 0,\ \mathbf{z}\in \mathbb{R}^n.$
\State$\mathbf{z}' = \varphi(\mathbf{z}; \tau)$
\State$\mathcal{S}_{\mathrm{eff}} = \mathcal{S}(\mathbf{z}') - \log \det J$
\For{$i\in [0,N)\cap \mathbb{Z}$}
\State $\mathbf{z}_{\mathrm{next}} = \mathbf{z}+\delta \mathbf{z},$ $\delta \mathbf{z}$ random, symmetric
\State $\mathbf{z}'_{\mathrm{next}} = \varphi(\mathbf{z}_\mathrm{next}; \tau)$
\State $\mathcal{S}_{\mathrm{eff},\mathrm{next}} = \mathcal{S}(\mathbf{z}'_{\mathrm{next}}) - \log \det J_{\mathrm{next}}$
\If{$\mathrm{Uniform}(0,1)\leq e^{-\mathrm{Re\,} (\mathcal{S}_{\mathrm{eff},\mathrm{next}}-\mathcal{S}_\mathrm{eff})}$}
\State $\mathbf{z} = \mathbf{z}_{\mathrm{next}}$
\State $\mathbf{z}' = \mathbf{z}'_{\mathrm{next}}$
\State $\mathcal{S}_{\mathrm{eff}} = \mathcal{S}_{\mathrm{eff},\mathrm{next}}$
\EndIf
\State Record $\mathbf{z},\mathbf{z}',\mathcal{S}_{\mathrm{eff}}.$
\EndFor
\end{algorithmic}
\end{algorithm}
\end{figure}

See Algorithm~\ref{alg:LTMC} for a pseudocode sketch of the approach. The algorithm is a Metropolis-Hastings algorithm, with samples taken on the initial integration manifold $\mathbb{R}^n$ according to the probability distribution $e^{-\mathcal{S}_{\mathrm{eff}}}$ on the thimbles, taking into account the requisite change of variables. The larger $\tau$ is, the more the probability landscape on the initial manifold will localize to small region(s) with high probability density.
As a result, for large $\tau$, the algorithm may become less effective at finding all relevant thimbles and arriving at the desired distribution.
As such, there emerge conflicting incentives to both minimize and maximize $\tau$, which implies there will be an optimal $\tau$ balancing these desired outcomes.

It should also be noted that calculating $\det J$ is by far the most computationally expensive aspect of this algorithm. In this work, we take the approach of simply numerically integrating Eq.~(\ref{eq:Jacobian_flow}), but more efficient schemes exist in the literature~\cite{alexandru2017schwinger, Cristoforetti_2014,Alexandru_estimator}.

A perennial issue in Monte Carlo methods is balancing the need to sufficiently explore the sample space and to spend time in high-weight regions of it. In the limit of infinite runtime this issue takes care of itself, but of course, no simulation has infinite time. Introduced in Ref.~\cite{DUANE1987216}, Hybrid (or Hamiltonian) Monte Carlo (HMC) provides an alternative sampling scheme designed to address this problem by providing a mechanism to more efficiently explore the parameter space than standard approaches. This approach augments the ``spatial'' coordinates $\mathbf{z}$ with an auxiliary momentum $\boldsymbol{\pi}\in \mathbb{R}^n$. This new set of variables are then flowed via Hamilton's equations with the Hamiltonian $\mathcal{H}(\mathbf{z}, \boldsymbol{\pi})= \mathcal{S}(\mathbf{z})+\frac{\boldsymbol{\pi}^2}{2}$:
\begin{align}
    \frac{\mathrm{d}z_i}{\mathrm{d}\tau} = \frac{\partial \mathcal{H}}{\partial \pi^i} &= \pi^i\\
    \frac{\mathrm{d}\pi^i}{\mathrm{d}\tau} = -\frac{\partial \mathcal{H}}{\partial z_i} &= -\frac{\partial \mathcal{S}}{\partial z_i}.
\end{align}

Define $F(\mathbf{z}, \boldsymbol{\pi}; \tau):= (\mathbf{z}(\tau), \boldsymbol{\pi}(\tau))$ to be the Hamiltonian flow of $(\mathbf{z}, \boldsymbol{\pi}).$ A detailed implementation of the HMC method for thimble models is given in Algorithm~\ref{alg:LTHMC}. 

\begin{figure}
\begin{algorithm}[H]
\caption{Generalized thimble Hybrid Monte Carlo}\label{alg:LTHMC}
\begin{algorithmic}[1]
\Require $N,\tau_1, \tau_2\geq 0,\ \mathbf{z}\in \mathbb{R}^n.$
\State$\mathbf{z}' = \varphi(\mathbf{z}; \tau)$
\State$\mathcal{S}_{\mathrm{eff}} = \mathcal{S}(\mathbf{z}') - \log \det J$
\For{$i\in [0,N)\cap \mathbb{Z}$}
\State $\boldsymbol{\pi}=\mathrm{Normal}(0,1)\in \mathbb{R}^n$
\State $(\mathbf{z}_{\mathrm{next}}, \boldsymbol{\pi}_{\mathrm{next}}) = F(\mathbf{z}, \boldsymbol{\pi};\tau_1)$
\State $\mathbf{z}'_{\mathrm{next}} = \varphi(\mathbf{z}_\mathrm{next}; \tau_2)$
\State $\mathcal{S}_{\mathrm{eff},\mathrm{next}} = \mathcal{S}(\mathbf{z}'_{\mathrm{next}}) - \log \det J_{\mathrm{next}}$
\If{$\mathrm{Uniform}(0,1)\leq e^{-\mathrm{Re\,} (\mathcal{S}_{\mathrm{eff},\mathrm{next}}-\mathcal{S}_\mathrm{eff}+\frac{\pi_{\mathrm{next}}^2-\pi^2}{2})}$}
\State $\mathbf{z} = \mathbf{z}_{\mathrm{next}}$
\State $\mathbf{z}' = \mathbf{z}'_{\mathrm{next}}$
\State $\mathcal{S}_{\mathrm{eff}} = \mathcal{S}_{\mathrm{eff},\mathrm{next}}$
\EndIf
\State Record $\mathbf{z},\mathbf{z}',\mathcal{S}_{\mathrm{eff}}.$
\EndFor
\end{algorithmic}
\end{algorithm}
\end{figure}

\section{Spin Coherent State Path Integrals}
While Lefschetz thimble Monte Carlo has been shown to be effective in ameliorating the sign problem, it only works with partition functions expressed as integrals, not the sums seen in standard path integral QMC for spin systems. To apply Lefschetz thimble Monte Carlo to spin systems, we need to write the spin partition function as an actual integral. This can be accomplished using a resolution of the identity expressible as an integral. Spin coherent states provide one such resolution. The novel application of Lefschetz thimble Monte Carlo to spin systems in a general way via spin coherent states is a key result of this paper.

We define a generic spin-$S$ spin coherent state as
\begin{equation}
    \ket{\mu} \equiv \frac{1}{(1+\vert\mu\vert^2)^S} \exp{\{\mu \hat{S}_-\}}\ket{\uparrow},
\end{equation}
where $\hat{S}_-\equiv \hat{S}_x-i\hat{S}_y$ is the lowering operator, $\ket{\uparrow}$ is the $+S$ state in the $z$-direction, and $
\mu\in\mathbb{C}$. Let $\mu = e^{i\varphi}\tan\frac{\theta}{2}$; then $\ket{\mu}$ corresponds to the +S eigenstate of the spin operator along an axis rotated from $+z$ by $\theta$ about the $y$-axis and then $\varphi$ about the $z$-axis \cite{radcliffe_properties_1971}.
Importantly, spin coherent states can resolve the identity as
\begin{equation}\label{eq:resI}
    I = \frac{2S+1}{\pi}\int_{\mathbb{R}^2} \frac{\mathrm{d}^2\mu}{\left(1+\vert\mu\vert^2\right)^2}\ket{\mu}\bra{\mu},
\end{equation}
where, for brevity, we let $\mathrm{d}^2\mu := \mathrm{d}(\mathrm{Re}\,\mu) \mathrm{d}(\mathrm{Im}\,\mu)$.

Let $\mu_j=x_j+iy_j$ at each ``imaginary time-slice'' $j\in \{0\cdots T-1\}$ of the usual path integral. 
Inserting Eq.~(\ref{eq:resI}) at each time-slice, we obtain the discrete \textit{spin coherent state path integral} up to $\mathcal{O}\big(\frac{\beta}{T}\big)$
\begin{equation}
        \label{eq:scpi1}
        \mathcal{Z} = \left(\prod_{j=0}^{T-1}\frac{2S+1}{\pi}\int_{\mathbb{R}^2} \frac{\mathrm{d}x_j\mathrm{d}y_j}{\left(1+ x_j^2+y_j^2\right)^2}\right)e^{-\mathcal{S}[\{x_j\},\{y_j\}]}
\end{equation}
where
\begin{align}
        \label{eq:scpi}
        \mathcal{S}[\{x_j\},\{y_j\}] \equiv\! 
        \sum_{j=0}^{T-1}\! \bigg(&\!2iS\frac{ y_{j+1}x_j-x_{j+1}y_j}{(1+x_j^2+y_j^2)}
        +\frac{\beta}{T}H^{cl}(x_j,y_j)\bigg).
\end{align}
and $H^{\mathrm{cl}}: \mathbb{R}^2\to \mathbb{R}$ is the so-called \textit{classical Hamiltonian} whose precise form depends on the problem of interest \cite{kochetov19952}. We refer the reader to Appendix~\ref{section:appA} for the details of this standard calculation. We can absorb the volume element of the integral in Eq.~(\ref{eq:scpi1}) into the action by defining $\mathcal{S}'= \mathcal{S} + 2\sum_{i}\log(1+x_i^2+y_i^2)$. 

Note the following regarding Eq.~(\ref{eq:scpi}): (a) The first term, a geometric phase, introduces its own sign problem into the partition function. In standard spin coherent state QMC \cite{takano1985monte}, one doesn't attempt to correct this sign problem or any original sign problem contained in $H^{\mathrm{cl}}$, potentially at exponential cost to the simulation algorithm; (b) Eq.~(\ref{eq:scpi}) requires the assumption that $y_{j+1}-y_j$ and $x_{j+1}-x_j$ are $\mathcal{O}\big(\frac{\beta}{T}\big).$ This is not mathematically well-founded at low spins \cite{shibata_note_1999, auerbach}, but can be somewhat justified at higher spins \cite{garg_spin_2001}. This is the uncontrolled approximation in the spin coherent path integral described in the introduction. We shall see that our results cast doubt onto the extent to which the approximation is reasonable, even for large spins, motivating a more rigorous understanding of this issue than provided by standard treatments of the spin coherent path integral;
(c) Both the geometric phase, and, it will turn out, the classical Hamiltonian, have a singularity when $x_j^2+y_j^2=-1.$ This causes the action to diverge in finite flow time. 
Such divergences are also observed when applying the Lefshetz thimble technique to fermionic models \cite{kanazawa2015structure, fujii2015lefschetz}. Consequently, when flowing the integration manifold, we seek a flow time sufficiently long to mitigate the sign problem, but not so long as to cause numerical blow ups. While unneccessary in the examples we consider, one can also avoid such numerical issues by modifying the flow so that it slows as one approaches singularities~\cite{tanizaki_gradient_2017}.

\section{Numerical Results}
Two systems are studied in this paper: a single spin-40 particle and a frustrated triplet of spin-10 particles. A combined HMC/holomorphic gradient flow algorithm was used for both systems. First, proposals are generated on the parameterization manifold using molecular dynamics evolution. We use the leapfrog integrator defined in~\cite{DUANE1987216} to implement this evolution. Once a proposal is generated, both initial and proposed points are flowed according to Eq.~(\ref{eq:flow}), and their Jacobians according to Eq.~(\ref{eq:Jacobian_flow}). The proposal is then accepted/rejected according step 8 of Alg.~\ref{alg:LTHMC}~\cite{code}. 

\subsection{Single Spin}
For the single spin-40 particle we consider the Hamiltonian $\hat{H}=\hat{S}_y$. This Hamiltonian has vanishing geometric phase, and thus has no sign problem. However, it serves as a good test to see whether the sign problem introduced by the spin coherent state path integral can be overcome by Lefschetz Spin QMC.

For this system we perform several studies. We first fix $T = 3$ and perform three sets of simulations: standard (unflowed) simulations sweeping across $\beta$, a sweep across $\tau$ at fixed beta, then finally a sweep across $\beta$ at fixed $\tau$. Unflowed simulations provide reference data for this proof-of-principles study, while a sweep across $\tau$ at fixed $\beta$ gives a rough idea as to what flow time is needed to handle the sign problem. Finally, with a suitable $\tau$ established, a sweep over $\beta$ can be performed. Unflowed simulations include $1 \times 10^6$ HMC steps, with proposals using a leapfrog integration of ``time" 1.0. The step-size of the integrator is chosen such that the acceptance rate remained above $90\%$ for all simulations. Such large statistics (on the scale of lattice QCD and lattice effective field theory simulations) are required to overcome the sign problem---even then, at small $\beta$ there is approximately $100\%$ uncertainty in observables.

\begin{figure}
    \centering
    \includegraphics[width=\columnwidth]{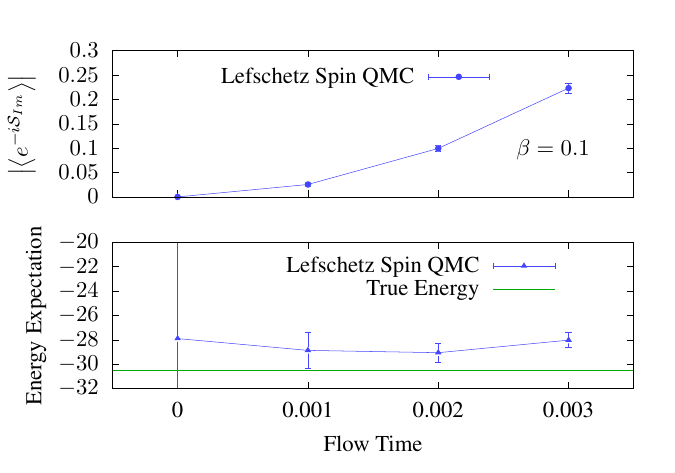}
    \caption{The expected sign (top) and energy expectation value (bottom) as a function of flow time $\tau$ for Lefschetz Spin QMC for the single spin example. Observe that while the flow is successful at mitgating the sign problem and the expectation value of energy is unchanged as we increase the flow time, there is a systematic error with respect to the true energy (obtained via exact diagonalization). }
    \label{fig:single_flow_combined}
\end{figure}

Flowed simulations take $5 \times 10^4$ steps for each value of $\beta$ and $\tau$ considered, and the HMC trajectory length is reduced to 0.05. This reduction is required because the deformation of the parameterization manifold is large enough that an HMC trajectory of length 1.0 results in essentially no acceptances. For the study of how the results depend on flow time we fix $\beta=0.1$, where the sign problem is the worst. From Fig.~\ref{fig:single_flow_combined}, it is clear that both the sign problem is mitigated and that the flow does not alter the value of observables. We find $\tau=0.003$ sufficient to tame the sign problem, so we use this flow time in our sweep across $\beta$.

\begin{figure}
    \centering
    \includegraphics[width=\columnwidth]{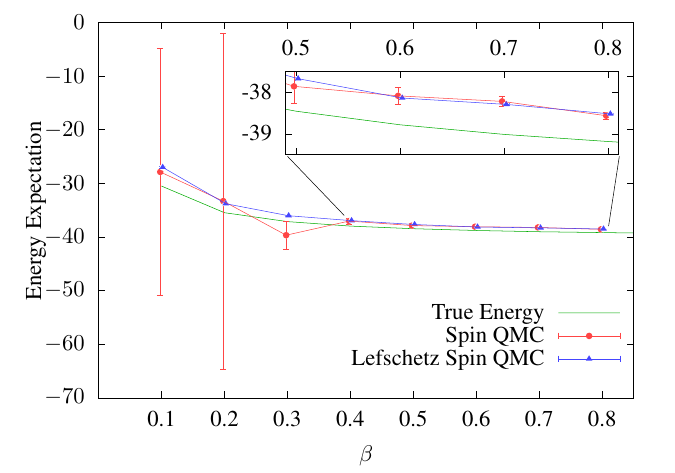}
    \caption{The expectation value of energy as a function of $\beta$ for the single spin example as computed via exact diagonalization, (unflowed) Spin QMC, and Lefschetz Spin QMC flowed for $\tau=0.003$.}
    \label{fig:single_energy}
\end{figure}

For this single spin system, each flowed step takes about 20 times longer than an unflowed step, and we have chosen the statistics of the flowed simulations to be such that the wall-clock time of both flowed and unflowed simulations are equal. Looking at Fig.~\ref{fig:single_energy}, it is clear that the statistical uncertainty of the flowed simulations are much smaller than unflowed simulations, especially at small $\beta$; since the two simulations take equal time, flowing results in demonstrable improvement in Monte Carlo performance.

\begin{figure}
    \centering
    \includegraphics[width=\columnwidth]{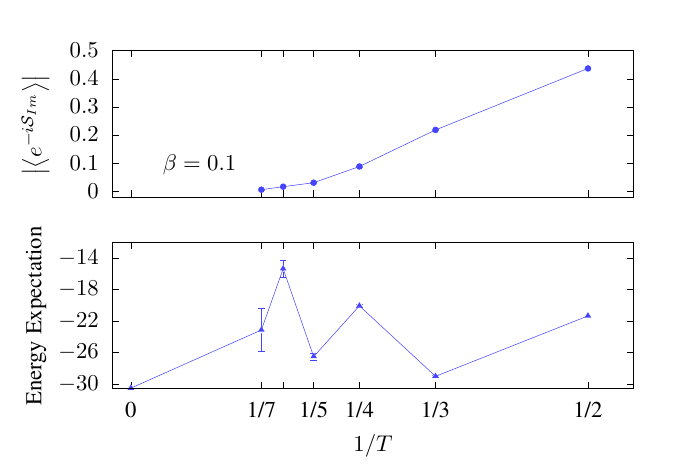}
    \caption{Average sign (top) and energy expectation value (bottom) for the single spin example at $\beta=0.1$ as a function of inverse time steps $\frac{1}{T}$ in the spin coherent path integral. We use $\tau=0.003$. Importantly, there is no evidence of a clear continuum limit as $\frac{1}{T}\rightarrow 0$ and we observe a staggered pattern, indicating odd/even dependence of the results. This provides numerical evidence of the breakdown of the spin coherent path integral due to the uncontrolled approximation that $|y_{j+1}-y_j|\sim|x_{j+1}-x_j|\sim \mathcal{O}\left(\frac{\beta}{T}\right)$ used in Eq.~(\ref{eq:scpi}).}
    \label{fig:single_cont_combined}
\end{figure}

Finally for the single spin system, we perform a time-continuum limit extrapolation at $\beta=0.1$, where the sign problem is severe. It is clear that at finite $T$ there is a sizeable difference between the path-integral and exact results. Furthermore this difference is largest at small $\beta$ as can be seen in Fig.~\ref{fig:single_cont_combined}. The question we wish to address is whether this difference goes to zero in the time continuum limit. If so, this discretization is demonstrably incorrect, in spite of its widespread usage in demonstrating the quantization of spin to half-integer values and textbook applications. That this discretization of the spin path integral contains uncontrolled errors, potentially leading to erroneous results, has been previously discussed in Refs.~\cite{auerbach, shibata_note_1999, galitski2011breakdown, kordas2016coherent, ranccon2020hubbard}, although it appears this is the first numerical demonstration of this fact, even for a single spin. The reason for this is the sign problem: for $T=3$ and certainly higher $T$'s, standard Monte Carlo simulations simply cannot resolve the phase oscillations in any reasonable time.

We simulate at $T\in\{2,3,\cdots,7\}$ to study the time continuum limit. Due to the severity of the sign problem at higher $T$, we fix $\tau=0.003$ and take $2.5\times 10^5$ steps in each simulation. We find a clear signal for the energy at every $T$, though the quality decreases at large $T$. Our results are displayed in Fig.~\ref{fig:single_cont_combined}. We find no evidence of a time continuum limit. Rather than smooth convergence to the continuum, a staggered pattern appears between even and odd number of timeslices. Furthermore, both branches trend upward, away from the correct result, as the continuum is approached. While it is logically possible that at extremely large $T$ the lattice results converge to the exact result, we find no compelling reason to believe this. We posit that, if a time continuum limit exists for this discretization, it does not converge to the correct value. We wish to reiterate that, though this is an extremely simple model, to our knowledge, this is the first direct numerical demonstration that this textbook discretization most likely does not converge to the correct continuum limit. 

\subsection{Frustrated Spin Triplet}
We also consider a frustrated spin triplet of three spin-10 particles interacting via the Hamiltonian $\hat{H} = \hat{S}_{z,1}\hat{S}_{z,2} + \hat{S}_{z,2}\hat{S}_{z,3} + \hat{S}_{z,3}\hat{S}_{z,1} + \hat{S}_{x,1}\hat{S}_{x,2} + \hat{S}_{x,2}\hat{S}_{x,3} + \hat{S}_{x,3}\hat{S}_{x,1}$. This example has a genuine sign problem~\cite{hen_resolution_2019}. Here, we again study $T=3$ and perform three classes of simulations: a sweep across $\beta$ on unflowed manifolds, a sweep across $\tau$ at fixed $\beta$, and finally a sweep across $\beta$ at fixed $\tau$. On unflowed manifolds we simulate $100\times 10^6$ HMC steps. Such high statistics are needed because the the sign problem of this discretization is severe. Even with these high statistics, we only barely find a signal at the large $\beta$. 

\begin{figure}
    \centering
    \includegraphics[width=\columnwidth]{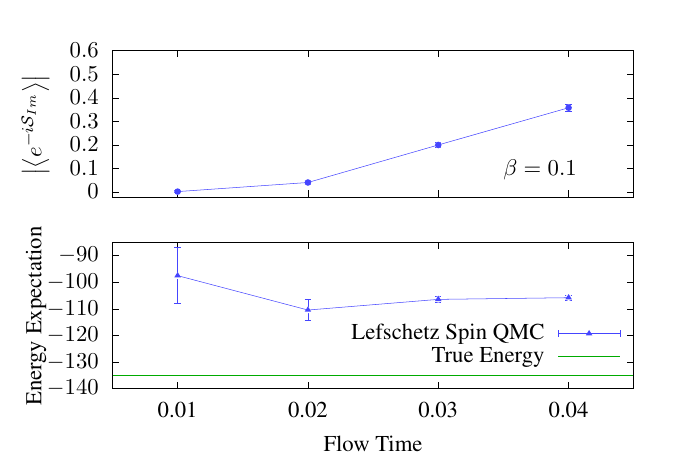}
    \caption{Average sign (top) and energy expectation value (bottom) for the spin triplet example. As in the single spin example, while the flow is successful at mitgating the sign problem, there is a systematic error with respect to the true energy.}
    \label{fig:triplet_flow_combined}
\end{figure}

\begin{figure}
    \centering
    \includegraphics[width=\columnwidth]{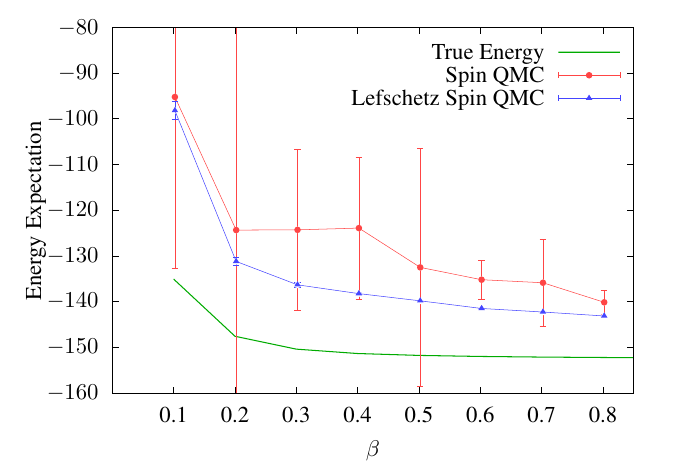}
    \caption{The expectation value of energy as a function of $\beta$ for the spin triplet example as computed via exact diagonalization, (unflowed) Spin QMC, and Lefschetz Spin QMC flowed for $\tau=0.015$.}
    \label{fig:triplet_energy}
\end{figure}

As in the single spin case, we first sweep across $\tau$ at fixed $\beta = 0.1$ to both demonstrate the correctness of the method and to establish a sufficient flow to tame the sign problem. These simulations consist of $1.25 \times 10^5$ steps and the results are plotted in Fig.~\ref{fig:triplet_flow_combined}. The statistical uncertainty shrinks with $\tau$ with no detectable deviation in the mean, demonstrating both the utility and correctness of the method.
To sweep across $\beta$, we fix $\tau=0.015$ and perform simulations with $1\times 10^6$ steps. 
While the unflowed data is extremely noisy, we find agreement between the two methods for all $\beta$, again demonstrating the correctness of the method. See Fig.~\ref{fig:triplet_energy}.

We conclude with a cost comparison: every flowed step of the triplet system costs 50 times more than an unflowed step. Therefore, twice as much wall-clock time is taken in the unflowed simulation than in the flowed simulation. However, the cheaper, flowed, data is drastically more precise. This again demonstrates that flowed simulations result in concrete gains.

\section{Conclusion and Outlook}
These results indicate that the generalized Lefschetz thimble method is successful at mitigating the sign problem in spin systems. These numerical examples served mostly as a proof of principle of these techniques, but nothing stops them from being applicable to larger problems. However, despite the demonstrated success of these techniques for mitigating the sign problem, we showed that uncontrolled approximations made in the standard spin coherent state path integral can introduce problematic systematic errors. While the possibility of such a breakdown in the spin coherent path integral in the continuous time limit was known, to our knowledge, our results provide the first numerical demonstration of such a breakdown. The reason for the lack of numerical demonstration is the sign problem: spin coherent path integrals introduce their own sign problem beyond any that may or may not be present in the initial spin Hamiltonian. Lefschetz thimbles allow us to overcome this sign problem. These systematic errors persist over all flows used, as well as over multiple choices for the number of time steps $T$ in the spin triplet path integral. To confidently apply Lefschetz thimble techniques to larger spin systems where the systematic error cannot be reliably evaluated will require addressing this issue with the standard spin coherent path integral. Our results indicate that this will be worthwhile, allowing us to unlock the potential of Lefschetz thimble methods for studying spin systems with a sign problem. 

\acknowledgements
This research was supported in part by the Heising-Simons Foundation, the Simons Foundation, and National Science Foundation Grant No. NSF PHY-1748958. T.C.M. and J.B.~acknowledge funding by the DoE ASCR Accelerated Research in Quantum Computing program (award No.~DE-SC0020312), DoE QSA, NSF QLCI (award No.~OMA-2120757), NSF PFCQC program, the DoE ASCR Quantum Testbed Pathfinder program (award No.~DE-SC0019040), AFOSR, ARO MURI, AFOSR MURI, and DARPA SAVaNT ADVENT. T.C.M. was partially supported by the National Institute of Standards and Technology (NIST) Summer Undergraduate Research Fellowship (SURF) program in the Information Technology Lab (ITL). J.B.~ was partially supported by the U.S.~Department of Energy, Office of Science, Office of Advanced Scientific Computing Research, Department of Energy Computational Science Graduate Fellowship (award No.~DE-SC0019323). N. C. W. was supported by the U.S. DOE under Grant No. DE-FG02- 00ER41132.  L.~T.~B. is a KBR employee working under the Prime Contract No. 80ARC020D0010 with the NASA Ames Research Center and is grateful for support from the DARPA Quantum Benchmarking program under IAA 8839, Annex 130. The United States Government retains, and by accepting the article for publication, the publisher acknowledges that the United States Government retains, a nonexclusive, paid-up, irrevocable, worldwide license to publish or reproduce the published form of this work, or allow others to do so, for United States Government purposes.

\bibliography{main.bib}

\begin{thebibliography}{56}%
\makeatletter
\providecommand \@ifxundefined [1]{%
 \@ifx{#1\undefined}
}%
\providecommand \@ifnum [1]{%
 \ifnum #1\expandafter \@firstoftwo
 \else \expandafter \@secondoftwo
 \fi
}%
\providecommand \@ifx [1]{%
 \ifx #1\expandafter \@firstoftwo
 \else \expandafter \@secondoftwo
 \fi
}%
\providecommand \natexlab [1]{#1}%
\providecommand \enquote  [1]{``#1''}%
\providecommand \bibnamefont  [1]{#1}%
\providecommand \bibfnamefont [1]{#1}%
\providecommand \citenamefont [1]{#1}%
\providecommand \href@noop [0]{\@secondoftwo}%
\providecommand \href [0]{\begingroup \@sanitize@url \@href}%
\providecommand \@href[1]{\@@startlink{#1}\@@href}%
\providecommand \@@href[1]{\endgroup#1\@@endlink}%
\providecommand \@sanitize@url [0]{\catcode `\\12\catcode `\$12\catcode
  `\&12\catcode `\#12\catcode `\^12\catcode `\_12\catcode `\%12\relax}%
\providecommand \@@startlink[1]{}%
\providecommand \@@endlink[0]{}%
\providecommand \url  [0]{\begingroup\@sanitize@url \@url }%
\providecommand \@url [1]{\endgroup\@href {#1}{\urlprefix }}%
\providecommand \urlprefix  [0]{URL }%
\providecommand \Eprint [0]{\href }%
\providecommand \doibase [0]{https://doi.org/}%
\providecommand \selectlanguage [0]{\@gobble}%
\providecommand \bibinfo  [0]{\@secondoftwo}%
\providecommand \bibfield  [0]{\@secondoftwo}%
\providecommand \translation [1]{[#1]}%
\providecommand \BibitemOpen [0]{}%
\providecommand \bibitemStop [0]{}%
\providecommand \bibitemNoStop [0]{.\EOS\space}%
\providecommand \EOS [0]{\spacefactor3000\relax}%
\providecommand \BibitemShut  [1]{\csname bibitem#1\endcsname}%
\let\auto@bib@innerbib\@empty
\bibitem [{\citenamefont {Troyer}\ and\ \citenamefont
  {Wiese}(2005)}]{troyer_computational_2005}%
  \BibitemOpen
  \bibfield  {author} {\bibinfo {author} {\bibfnamefont {M.}~\bibnamefont
  {Troyer}}\ and\ \bibinfo {author} {\bibfnamefont {U.-J.}\ \bibnamefont
  {Wiese}},\ }\bibfield  {title} {\bibinfo {title} {Computational complexity
  and fundamental limitations to fermionic quantum {Monte} {Carlo}
  simulations},\ }\href {https://doi.org/10.1103/PhysRevLett.94.170201}
  {\bibfield  {journal} {\bibinfo  {journal} {Phys. Rev. Lett.}\ }\textbf
  {\bibinfo {volume} {94}},\ \bibinfo {pages} {170201} (\bibinfo {year}
  {2005})}\BibitemShut {NoStop}%
\bibitem [{\citenamefont {Marvian}\ \emph {et~al.}(2019)\citenamefont
  {Marvian}, \citenamefont {Lidar},\ and\ \citenamefont
  {Hen}}]{marvian_computational_2019}%
  \BibitemOpen
  \bibfield  {author} {\bibinfo {author} {\bibfnamefont {M.}~\bibnamefont
  {Marvian}}, \bibinfo {author} {\bibfnamefont {D.~A.}\ \bibnamefont {Lidar}},\
  and\ \bibinfo {author} {\bibfnamefont {I.}~\bibnamefont {Hen}},\ }\bibfield
  {title} {\bibinfo {title} {On the computational complexity of curing
  non-stoquastic {{H}amiltonians}},\ }\href
  {https://doi.org/10.1038/s41467-019-09501-6} {\bibfield  {journal} {\bibinfo
  {journal} {Nat Commun}\ }\textbf {\bibinfo {volume} {10}},\ \bibinfo {pages}
  {1571} (\bibinfo {year} {2019})}\BibitemShut {NoStop}%
\bibitem [{\citenamefont {Hen}(2019)}]{hen_resolution_2019}%
  \BibitemOpen
  \bibfield  {author} {\bibinfo {author} {\bibfnamefont {I.}~\bibnamefont
  {Hen}},\ }\bibfield  {title} {\bibinfo {title} {Resolution of the sign
  problem for a frustrated triplet of spins},\ }\href
  {http://doi.org/abs/10.1103/PhysRevE.99.033306} {\bibfield  {journal}
  {\bibinfo  {journal} {Phys. Rev. E}\ }\textbf {\bibinfo {volume} {99}},\
  \bibinfo {pages} {033306} (\bibinfo {year} {2019})}\BibitemShut {NoStop}%
\bibitem [{\citenamefont {Li}\ \emph {et~al.}(2015)\citenamefont {Li},
  \citenamefont {Jiang},\ and\ \citenamefont {Yao}}]{Li_2015}%
  \BibitemOpen
  \bibfield  {author} {\bibinfo {author} {\bibfnamefont {Z.-X.}\ \bibnamefont
  {Li}}, \bibinfo {author} {\bibfnamefont {Y.-F.}\ \bibnamefont {Jiang}},\ and\
  \bibinfo {author} {\bibfnamefont {H.}~\bibnamefont {Yao}},\ }\bibfield
  {title} {\bibinfo {title} {Solving the fermion sign problem in quantum {Monte
  Carlo} simulations by {Majorana} representation},\ }\href
  {http://dx.doi.org/10.1103/PhysRevB.91.241117} {\bibfield  {journal}
  {\bibinfo  {journal} {Phys. Rev. B}\ }\textbf {\bibinfo {volume} {91}},\
  \bibinfo {pages} {241117} (\bibinfo {year} {2015})}\BibitemShut {NoStop}%
\bibitem [{\citenamefont {Wan}\ \emph {et~al.}(2020)\citenamefont {Wan},
  \citenamefont {Zhang},\ and\ \citenamefont {Yao}}]{wan2020mitigating}%
  \BibitemOpen
  \bibfield  {author} {\bibinfo {author} {\bibfnamefont {Z.-Q.}\ \bibnamefont
  {Wan}}, \bibinfo {author} {\bibfnamefont {S.-X.}\ \bibnamefont {Zhang}},\
  and\ \bibinfo {author} {\bibfnamefont {H.}~\bibnamefont {Yao}},\ }\bibfield
  {title} {\bibinfo {title} {Mitigating sign problem by automatic
  differentiation},\ }\href {https://arxiv.org/abs/2010.01141} {\bibfield
  {journal} {\bibinfo  {journal} {arXiv: 2010.01141}\ } (\bibinfo {year}
  {2020})}\BibitemShut {NoStop}%
\bibitem [{\citenamefont {Hangleiter}\ \emph {et~al.}(2020)\citenamefont
  {Hangleiter}, \citenamefont {Roth}, \citenamefont {Nagaj},\ and\
  \citenamefont {Eisert}}]{hangleiter_easing_2020}%
  \BibitemOpen
  \bibfield  {author} {\bibinfo {author} {\bibfnamefont {D.}~\bibnamefont
  {Hangleiter}}, \bibinfo {author} {\bibfnamefont {I.}~\bibnamefont {Roth}},
  \bibinfo {author} {\bibfnamefont {D.}~\bibnamefont {Nagaj}},\ and\ \bibinfo
  {author} {\bibfnamefont {J.}~\bibnamefont {Eisert}},\ }\bibfield  {title}
  {\bibinfo {title} {Easing the {Monte} {Carlo} sign problem},\ }\href
  {https://dx.doi.org/10.1126/sciadv.abb8341} {\bibfield  {journal} {\bibinfo
  {journal} {Sci. Adv.}\ }\textbf {\bibinfo {volume} {6}},\ \bibinfo {pages}
  {eabb8341} (\bibinfo {year} {2020})}\BibitemShut {NoStop}%
\bibitem [{\citenamefont {Bravyi}\ \emph
  {et~al.}(2006{\natexlab{a}})\citenamefont {Bravyi}, \citenamefont
  {Divincenzo}, \citenamefont {Oliveira},\ and\ \citenamefont
  {Terhal}}]{bravyi2006complexity}%
  \BibitemOpen
  \bibfield  {author} {\bibinfo {author} {\bibfnamefont {S.}~\bibnamefont
  {Bravyi}}, \bibinfo {author} {\bibfnamefont {D.~P.}\ \bibnamefont
  {Divincenzo}}, \bibinfo {author} {\bibfnamefont {R.~I.}\ \bibnamefont
  {Oliveira}},\ and\ \bibinfo {author} {\bibfnamefont {B.~M.}\ \bibnamefont
  {Terhal}},\ }\bibfield  {title} {\bibinfo {title} {The complexity of
  stoquastic local {H}amiltonian problems},\ }\href
  {https://arxiv.org/abs/quant-ph/0606140} {\bibfield  {journal} {\bibinfo
  {journal} {arXiv:quant-ph/0606140}\ } (\bibinfo {year}
  {2006}{\natexlab{a}})}\BibitemShut {NoStop}%
\bibitem [{\citenamefont {Jarret}(2018)}]{jarret2018hamiltonian}%
  \BibitemOpen
  \bibfield  {author} {\bibinfo {author} {\bibfnamefont {M.}~\bibnamefont
  {Jarret}},\ }\bibfield  {title} {\bibinfo {title} {Hamiltonian surgery:
  Cheeger-type gap inequalities for nonpositive (stoquastic), real, and
  hermitian matrices},\ }\href {https://arxiv.org/abs/1804.06857} {\bibfield
  {journal} {\bibinfo  {journal} {arXiv preprint arXiv:1804.06857}\ } (\bibinfo
  {year} {2018})}\BibitemShut {NoStop}%
\bibitem [{\citenamefont {Hen}(2021)}]{Hen_2021}%
  \BibitemOpen
  \bibfield  {author} {\bibinfo {author} {\bibfnamefont {I.}~\bibnamefont
  {Hen}},\ }\bibfield  {title} {\bibinfo {title} {Determining quantum {Monte
  Carlo} simulability with geometric phases},\ }\href
  {http://dx.doi.org/10.1103/PhysRevResearch.3.023080} {\bibfield  {journal}
  {\bibinfo  {journal} {Phys. Rev. R.}\ }\textbf {\bibinfo {volume} {3}},\
  \bibinfo {pages} {023080} (\bibinfo {year} {2021})}\BibitemShut {NoStop}%
\bibitem [{\citenamefont {Bravyi}\ and\ \citenamefont
  {Terhal}(2010)}]{bravyi2008complexity}%
  \BibitemOpen
  \bibfield  {author} {\bibinfo {author} {\bibfnamefont {S.}~\bibnamefont
  {Bravyi}}\ and\ \bibinfo {author} {\bibfnamefont {B.}~\bibnamefont
  {Terhal}},\ }\bibfield  {title} {\bibinfo {title} {Complexity of stoquastic
  frustration-free hamiltonians},\ }\href {https://doi.org/10.1137/08072689X}
  {\bibfield  {journal} {\bibinfo  {journal} {Siam J. Comput.}\ }\textbf
  {\bibinfo {volume} {39}},\ \bibinfo {pages} {1462} (\bibinfo {year}
  {2010})}\BibitemShut {NoStop}%
\bibitem [{\citenamefont {Bravyi}\ \emph
  {et~al.}(2006{\natexlab{b}})\citenamefont {Bravyi}, \citenamefont {Bessen},\
  and\ \citenamefont {Terhal}}]{bravyi2006MerlinArthurGA}%
  \BibitemOpen
  \bibfield  {author} {\bibinfo {author} {\bibfnamefont {S.}~\bibnamefont
  {Bravyi}}, \bibinfo {author} {\bibfnamefont {A.~J.}\ \bibnamefont {Bessen}},\
  and\ \bibinfo {author} {\bibfnamefont {B.}~\bibnamefont {Terhal}},\
  }\bibfield  {title} {\bibinfo {title} {Merlin-arthur games and stoquastic
  complexity},\ }\href {https://arxiv.org/abs/quant-ph/0606140} {\bibfield
  {journal} {\bibinfo  {journal} {arXiv:quant-ph/0606140}\ } (\bibinfo {year}
  {2006}{\natexlab{b}})}\BibitemShut {NoStop}%
\bibitem [{\citenamefont {Crosson}\ and\ \citenamefont
  {Harrow}(2021)}]{crosson_2021}%
  \BibitemOpen
  \bibfield  {author} {\bibinfo {author} {\bibfnamefont {E.}~\bibnamefont
  {Crosson}}\ and\ \bibinfo {author} {\bibfnamefont {A.}~\bibnamefont
  {Harrow}},\ }\bibfield  {title} {\bibinfo {title} {Rapid mixing of path
  integral monte carlo for 1d stoquastic hamiltonians},\ }\href
  {https://quantum-journal.org/papers/q-2021-02-11-395/} {\bibfield  {journal}
  {\bibinfo  {journal} {Quantum}\ }\textbf {\bibinfo {volume} {5}},\ \bibinfo
  {pages} {395} (\bibinfo {year} {2021})}\BibitemShut {NoStop}%
\bibitem [{\citenamefont {Bringewatt}\ and\ \citenamefont
  {Jarret}(2020)}]{Bringewatt_2020}%
  \BibitemOpen
  \bibfield  {author} {\bibinfo {author} {\bibfnamefont {J.}~\bibnamefont
  {Bringewatt}}\ and\ \bibinfo {author} {\bibfnamefont {M.}~\bibnamefont
  {Jarret}},\ }\bibfield  {title} {\bibinfo {title} {Effective gaps are not
  effective: Quasipolynomial classical simulation of obstructed stoquastic
  {H}amiltonians},\ }\href {http://dx.doi.org/10.1103/PhysRevLett.125.170504}
  {\bibfield  {journal} {\bibinfo  {journal} {Phys. Rev. Lett.}\ }\textbf
  {\bibinfo {volume} {125}},\ \bibinfo {pages} {170504} (\bibinfo {year}
  {2020})}\BibitemShut {NoStop}%
\bibitem [{\citenamefont {Hastings}(2020)}]{hastings2020power}%
  \BibitemOpen
  \bibfield  {author} {\bibinfo {author} {\bibfnamefont {M.~B.}\ \bibnamefont
  {Hastings}},\ }\bibfield  {title} {\bibinfo {title} {The power of adiabatic
  quantum computation with no sign problem},\ }\href
  {http://arxiv.org/abs/2005.03791} {\bibfield  {journal} {\bibinfo  {journal}
  {arXiv:2005.03791}\ } (\bibinfo {year} {2020})}\BibitemShut {NoStop}%
\bibitem [{\citenamefont {Aharonov}\ \emph {et~al.}(2008)\citenamefont
  {Aharonov}, \citenamefont {Van~Dam}, \citenamefont {Kempe}, \citenamefont
  {Landau}, \citenamefont {Lloyd},\ and\ \citenamefont
  {Regev}}]{aharonov2008adiabatic}%
  \BibitemOpen
  \bibfield  {author} {\bibinfo {author} {\bibfnamefont {D.}~\bibnamefont
  {Aharonov}}, \bibinfo {author} {\bibfnamefont {W.}~\bibnamefont {Van~Dam}},
  \bibinfo {author} {\bibfnamefont {J.}~\bibnamefont {Kempe}}, \bibinfo
  {author} {\bibfnamefont {Z.}~\bibnamefont {Landau}}, \bibinfo {author}
  {\bibfnamefont {S.}~\bibnamefont {Lloyd}},\ and\ \bibinfo {author}
  {\bibfnamefont {O.}~\bibnamefont {Regev}},\ }\bibfield  {title} {\bibinfo
  {title} {Adiabatic quantum computation is equivalent to standard quantum
  computation},\ }\href {https://doi.org/10.1137/080734479} {\bibfield
  {journal} {\bibinfo  {journal} {SIAM review}\ }\textbf {\bibinfo {volume}
  {50}},\ \bibinfo {pages} {755} (\bibinfo {year} {2008})}\BibitemShut
  {NoStop}%
\bibitem [{\citenamefont {Pham}(1983)}]{pham1983vanishing}%
  \BibitemOpen
  \bibfield  {author} {\bibinfo {author} {\bibfnamefont {F.}~\bibnamefont
  {Pham}},\ }\bibfield  {title} {\bibinfo {title} {Vanishing homologies and the
  $n$ variable saddlepoint method},\ }\href
  {https://www.ams.org/books/pspum/040.2/} {\bibfield  {journal} {\bibinfo
  {journal} {Proc. Symp. Pure Math.}\ }\textbf {\bibinfo {volume} {40}},\
  \bibinfo {pages} {310} (\bibinfo {year} {1983})}\BibitemShut {NoStop}%
\bibitem [{\citenamefont {Witten}(2010)}]{witten2010new}%
  \BibitemOpen
  \bibfield  {author} {\bibinfo {author} {\bibfnamefont {E.}~\bibnamefont
  {Witten}},\ }\bibfield  {title} {\bibinfo {title} {A new look at the path
  integral of quantum mechanics},\ }\href {https://arxiv.org/abs/1009.6032}
  {\bibfield  {journal} {\bibinfo  {journal} {Surv. Differ. Geom.}\ }\textbf
  {\bibinfo {volume} {15}},\ \bibinfo {pages} {345} (\bibinfo {year}
  {2010})}\BibitemShut {NoStop}%
\bibitem [{\citenamefont {Witten}(2011)}]{witten2011analytic}%
  \BibitemOpen
  \bibfield  {author} {\bibinfo {author} {\bibfnamefont {E.}~\bibnamefont
  {Witten}},\ }\bibfield  {title} {\bibinfo {title} {Analytic continuation of
  {C}hern-{S}imons theory},\ }\href {https://arxiv.org/abs/1001.2933}
  {\bibfield  {journal} {\bibinfo  {journal} {AMS/IP Stud. Adv. Math}\ }\textbf
  {\bibinfo {volume} {50}},\ \bibinfo {pages} {347} (\bibinfo {year}
  {2011})}\BibitemShut {NoStop}%
\bibitem [{\citenamefont {Cristoforetti}\ \emph {et~al.}(2012)\citenamefont
  {Cristoforetti}, \citenamefont {Di~Renzo},\ and\ \citenamefont
  {Scorzato}}]{Cristoforetti_2012}%
  \BibitemOpen
  \bibfield  {author} {\bibinfo {author} {\bibfnamefont {M.}~\bibnamefont
  {Cristoforetti}}, \bibinfo {author} {\bibfnamefont {F.}~\bibnamefont
  {Di~Renzo}},\ and\ \bibinfo {author} {\bibfnamefont {L.}~\bibnamefont
  {Scorzato}},\ }\bibfield  {title} {\bibinfo {title} {New approach to the sign
  problem in quantum field theories: High density {Q}{C}{D} on a {L}efschetz
  thimble},\ }\href {http://dx.doi.org/10.1103/PhysRevD.86.074506} {\bibfield
  {journal} {\bibinfo  {journal} {Phys. Rev. D}\ }\textbf {\bibinfo {volume}
  {86}},\ \bibinfo {pages} {074506} (\bibinfo {year} {2012})}\BibitemShut
  {NoStop}%
\bibitem [{\citenamefont {Cristoforetti}\ \emph {et~al.}(2013)\citenamefont
  {Cristoforetti}, \citenamefont {Di~Renzo}, \citenamefont {Mukherjee},\ and\
  \citenamefont {Scorzato}}]{cristoforetti2013montecarlo}%
  \BibitemOpen
  \bibfield  {author} {\bibinfo {author} {\bibfnamefont {M.}~\bibnamefont
  {Cristoforetti}}, \bibinfo {author} {\bibfnamefont {F.}~\bibnamefont
  {Di~Renzo}}, \bibinfo {author} {\bibfnamefont {A.}~\bibnamefont
  {Mukherjee}},\ and\ \bibinfo {author} {\bibfnamefont {L.}~\bibnamefont
  {Scorzato}},\ }\bibfield  {title} {\bibinfo {title} {Monte carlo simulations
  on the lefschetz thimble: Taming the sign problem},\ }\href
  {https://doi.org/10.1103/PhysRevD.88.051501} {\bibfield  {journal} {\bibinfo
  {journal} {Phys. Rev. D}\ }\textbf {\bibinfo {volume} {88}},\ \bibinfo
  {pages} {051501} (\bibinfo {year} {2013})}\BibitemShut {NoStop}%
\bibitem [{\citenamefont {Mukherjee}\ \emph {et~al.}(2013)\citenamefont
  {Mukherjee}, \citenamefont {Cristoforetti},\ and\ \citenamefont
  {Scorzato}}]{mukherjee2013metropolis}%
  \BibitemOpen
  \bibfield  {author} {\bibinfo {author} {\bibfnamefont {A.}~\bibnamefont
  {Mukherjee}}, \bibinfo {author} {\bibfnamefont {M.}~\bibnamefont
  {Cristoforetti}},\ and\ \bibinfo {author} {\bibfnamefont {L.}~\bibnamefont
  {Scorzato}},\ }\bibfield  {title} {\bibinfo {title} {Metropolis {M}onte
  {C}arlo integration on the {L}efschetz thimble: Application to a
  one-plaquette model},\ }\href {https://doi.org/10.1103/PhysRevD.88.051502}
  {\bibfield  {journal} {\bibinfo  {journal} {Phys. Rev. D}\ }\textbf {\bibinfo
  {volume} {88}},\ \bibinfo {pages} {051502} (\bibinfo {year}
  {2013})}\BibitemShut {NoStop}%
\bibitem [{\citenamefont {Fujii}\ \emph {et~al.}(2013)\citenamefont {Fujii},
  \citenamefont {Honda}, \citenamefont {Kato}, \citenamefont {Kikukawa},
  \citenamefont {Komatsu},\ and\ \citenamefont {Sano}}]{Fujii_2013}%
  \BibitemOpen
  \bibfield  {author} {\bibinfo {author} {\bibfnamefont {H.}~\bibnamefont
  {Fujii}}, \bibinfo {author} {\bibfnamefont {D.}~\bibnamefont {Honda}},
  \bibinfo {author} {\bibfnamefont {M.}~\bibnamefont {Kato}}, \bibinfo {author}
  {\bibfnamefont {Y.}~\bibnamefont {Kikukawa}}, \bibinfo {author}
  {\bibfnamefont {S.}~\bibnamefont {Komatsu}},\ and\ \bibinfo {author}
  {\bibfnamefont {T.}~\bibnamefont {Sano}},\ }\bibfield  {title} {\bibinfo
  {title} {Hybrid {Monte} {Carlo} on {L}efschetz thimbles — a study of the
  residual sign problem},\ }\href {http://dx.doi.org/10.1007/JHEP10(2013)147}
  {\bibfield  {journal} {\bibinfo  {journal} {J. High Energy Phys{.}}\ }\textbf
  {\bibinfo {volume} {2013}},\ \bibinfo {pages} {147} (\bibinfo {year}
  {2013})}\BibitemShut {NoStop}%
\bibitem [{\citenamefont {Aarts}(2013)}]{aarts2013lefschetz}%
  \BibitemOpen
  \bibfield  {author} {\bibinfo {author} {\bibfnamefont {G.}~\bibnamefont
  {Aarts}},\ }\bibfield  {title} {\bibinfo {title} {Lefschetz thimbles and
  stochastic quantization: Complex actions in the complex plane},\ }\href
  {https://doi.org/10.1103/PhysRevD.88.094501} {\bibfield  {journal} {\bibinfo
  {journal} {Phys. Rev. D}\ }\textbf {\bibinfo {volume} {88}},\ \bibinfo
  {pages} {094501} (\bibinfo {year} {2013})}\BibitemShut {NoStop}%
\bibitem [{\citenamefont {Cristoforetti}\ \emph {et~al.}(2014)\citenamefont
  {Cristoforetti}, \citenamefont {Di~Renzo}, \citenamefont {Eruzzi},
  \citenamefont {Mukherjee}, \citenamefont {Schmidt}, \citenamefont
  {Scorzato},\ and\ \citenamefont {Torrero}}]{Cristoforetti_2014}%
  \BibitemOpen
  \bibfield  {author} {\bibinfo {author} {\bibfnamefont {M.}~\bibnamefont
  {Cristoforetti}}, \bibinfo {author} {\bibfnamefont {F.}~\bibnamefont
  {Di~Renzo}}, \bibinfo {author} {\bibfnamefont {G.}~\bibnamefont {Eruzzi}},
  \bibinfo {author} {\bibfnamefont {A.}~\bibnamefont {Mukherjee}}, \bibinfo
  {author} {\bibfnamefont {C.}~\bibnamefont {Schmidt}}, \bibinfo {author}
  {\bibfnamefont {L.}~\bibnamefont {Scorzato}},\ and\ \bibinfo {author}
  {\bibfnamefont {C.}~\bibnamefont {Torrero}},\ }\bibfield  {title} {\bibinfo
  {title} {An efficient method to compute the residual phase on a {L}efschetz
  thimble},\ }\href {http://dx.doi.org/10.1103/PhysRevD.89.114505} {\bibfield
  {journal} {\bibinfo  {journal} {Phys. Rev. D}\ }\textbf {\bibinfo {volume}
  {89}},\ \bibinfo {pages} {114505} (\bibinfo {year} {2014})}\BibitemShut
  {NoStop}%
\bibitem [{\citenamefont {Aarts}\ \emph {et~al.}(2014)\citenamefont {Aarts},
  \citenamefont {Bongiovanni}, \citenamefont {Seiler},\ and\ \citenamefont
  {Sexty}}]{aarts2014some}%
  \BibitemOpen
  \bibfield  {author} {\bibinfo {author} {\bibfnamefont {G.}~\bibnamefont
  {Aarts}}, \bibinfo {author} {\bibfnamefont {L.}~\bibnamefont {Bongiovanni}},
  \bibinfo {author} {\bibfnamefont {E.}~\bibnamefont {Seiler}},\ and\ \bibinfo
  {author} {\bibfnamefont {D.}~\bibnamefont {Sexty}},\ }\bibfield  {title}
  {\bibinfo {title} {Some remarks on lefschetz thimbles and complex langevin
  dynamics},\ }\href {https://doi.org/10.1007/JHEP10(2014)159} {\bibfield
  {journal} {\bibinfo  {journal} {J. High Energ. Phys.}\ }\textbf {\bibinfo
  {volume} {2014}}\bibinfo  {number} { (10)},\ \bibinfo {pages}
  {1}}\BibitemShut {NoStop}%
\bibitem [{\citenamefont {Kanazawa}\ and\ \citenamefont
  {Tanizaki}(2015)}]{kanazawa2015structure}%
  \BibitemOpen
\bibfield  {number} {  }\bibfield  {author} {\bibinfo {author} {\bibfnamefont
  {T.}~\bibnamefont {Kanazawa}}\ and\ \bibinfo {author} {\bibfnamefont
  {Y.}~\bibnamefont {Tanizaki}},\ }\bibfield  {title} {\bibinfo {title}
  {Structure of lefschetz thimbles in simple fermionic systems},\ }\href
  {https://doi.org/10.1007/JHEP03(2015)044} {\bibfield  {journal} {\bibinfo
  {journal} {J. High Energ. Phys.}\ }\textbf {\bibinfo {volume} {2015}}\bibinfo
   {number} { (3)},\ \bibinfo {pages} {1}}\BibitemShut {NoStop}%
\bibitem [{\citenamefont {Fujii}\ \emph {et~al.}(2015)\citenamefont {Fujii},
  \citenamefont {Kamata},\ and\ \citenamefont {Kikukawa}}]{fujii2015lefschetz}%
  \BibitemOpen
\bibfield  {number} {  }\bibfield  {author} {\bibinfo {author} {\bibfnamefont
  {H.}~\bibnamefont {Fujii}}, \bibinfo {author} {\bibfnamefont
  {S.}~\bibnamefont {Kamata}},\ and\ \bibinfo {author} {\bibfnamefont
  {Y.}~\bibnamefont {Kikukawa}},\ }\bibfield  {title} {\bibinfo {title}
  {{L}efschetz thimble structure in one-dimensional lattice {T}hirring model at
  finite density},\ }\href {https://doi.org/10.1007/JHEP11(2015)078} {\bibfield
   {journal} {\bibinfo  {journal} {J. High Energy Phys{.}}\ }\textbf {\bibinfo
  {volume} {2015}},\ \bibinfo {pages} {1} (\bibinfo {year} {2015})}\BibitemShut
  {NoStop}%
\bibitem [{\citenamefont {Alexandru}\ \emph
  {et~al.}(2016{\natexlab{a}})\citenamefont {Alexandru}, \citenamefont
  {Ba\ifmmode~\mbox{\c{s}}\else \c{s}\fi{}ar},\ and\ \citenamefont
  {Bedaque}}]{alexandru2016mc}%
  \BibitemOpen
  \bibfield  {author} {\bibinfo {author} {\bibfnamefont {A.}~\bibnamefont
  {Alexandru}}, \bibinfo {author} {\bibfnamefont {G.}~\bibnamefont
  {Ba\ifmmode~\mbox{\c{s}}\else \c{s}\fi{}ar}},\ and\ \bibinfo {author}
  {\bibfnamefont {P.}~\bibnamefont {Bedaque}},\ }\bibfield  {title} {\bibinfo
  {title} {Monte {Carlo} algorithm for simulating fermions on {L}efschetz
  thimbles},\ }\href {https://doi.org/10.1103/PhysRevD.93.014504} {\bibfield
  {journal} {\bibinfo  {journal} {Phys. Rev. D}\ }\textbf {\bibinfo {volume}
  {93}},\ \bibinfo {pages} {014504} (\bibinfo {year}
  {2016}{\natexlab{a}})}\BibitemShut {NoStop}%
\bibitem [{\citenamefont {Alexandru}\ \emph
  {et~al.}(2017{\natexlab{a}})\citenamefont {Alexandru}, \citenamefont
  {Ba\ifmmode~\mbox{\c{s}}\else \c{s}\fi{}ar}, \citenamefont {Bedaque},
  \citenamefont {Ridgway},\ and\ \citenamefont {Warrington}}]{alexandru2017mc}%
  \BibitemOpen
  \bibfield  {author} {\bibinfo {author} {\bibfnamefont {A.}~\bibnamefont
  {Alexandru}}, \bibinfo {author} {\bibfnamefont {G.}~\bibnamefont
  {Ba\ifmmode~\mbox{\c{s}}\else \c{s}\fi{}ar}}, \bibinfo {author}
  {\bibfnamefont {P.~F.}\ \bibnamefont {Bedaque}}, \bibinfo {author}
  {\bibfnamefont {G.~W.}\ \bibnamefont {Ridgway}},\ and\ \bibinfo {author}
  {\bibfnamefont {N.~C.}\ \bibnamefont {Warrington}},\ }\bibfield  {title}
  {\bibinfo {title} {Monte {Carlo} calculations of the finite density
  {T}hirring model},\ }\href {https://doi.org/10.1103/PhysRevD.95.014502}
  {\bibfield  {journal} {\bibinfo  {journal} {Phys. Rev. D}\ }\textbf {\bibinfo
  {volume} {95}},\ \bibinfo {pages} {014502} (\bibinfo {year}
  {2017}{\natexlab{a}})}\BibitemShut {NoStop}%
\bibitem [{\citenamefont {Alexandru}\ \emph
  {et~al.}(2018{\natexlab{a}})\citenamefont {Alexandru}, \citenamefont
  {Bedaque}, \citenamefont {Lamm}, \citenamefont {Lawrence},\ and\
  \citenamefont {Warrington}}]{alexandru2018fermions}%
  \BibitemOpen
  \bibfield  {author} {\bibinfo {author} {\bibfnamefont {A.}~\bibnamefont
  {Alexandru}}, \bibinfo {author} {\bibfnamefont {P.~F.}\ \bibnamefont
  {Bedaque}}, \bibinfo {author} {\bibfnamefont {H.}~\bibnamefont {Lamm}},
  \bibinfo {author} {\bibfnamefont {S.}~\bibnamefont {Lawrence}},\ and\
  \bibinfo {author} {\bibfnamefont {N.~C.}\ \bibnamefont {Warrington}},\
  }\bibfield  {title} {\bibinfo {title} {Fermions at finite density in $2+1$
  dimensions with sign-optimized manifolds},\ }\href
  {https://doi.org/10.1103/PhysRevLett.121.191602} {\bibfield  {journal}
  {\bibinfo  {journal} {Phys. Rev. Lett.}\ }\textbf {\bibinfo {volume} {121}},\
  \bibinfo {pages} {191602} (\bibinfo {year} {2018}{\natexlab{a}})}\BibitemShut
  {NoStop}%
\bibitem [{\citenamefont {Di~Renzo}\ and\ \citenamefont
  {Eruzzi}(2018)}]{direnzo2018one}%
  \BibitemOpen
  \bibfield  {author} {\bibinfo {author} {\bibfnamefont {F.}~\bibnamefont
  {Di~Renzo}}\ and\ \bibinfo {author} {\bibfnamefont {G.}~\bibnamefont
  {Eruzzi}},\ }\bibfield  {title} {\bibinfo {title} {One-dimensional qcd in
  thimble regularization},\ }\href {https://doi.org/10.1103/PhysRevD.97.014503}
  {\bibfield  {journal} {\bibinfo  {journal} {Phys. Rev. D}\ }\textbf {\bibinfo
  {volume} {97}},\ \bibinfo {pages} {014503} (\bibinfo {year}
  {2018})}\BibitemShut {NoStop}%
\bibitem [{\citenamefont {Alexandru}\ \emph
  {et~al.}(2018{\natexlab{b}})\citenamefont {Alexandru}, \citenamefont
  {Ba\ifmmode~\mbox{\c{s}}\else \c{s}\fi{}ar}, \citenamefont {Bedaque},
  \citenamefont {Lamm},\ and\ \citenamefont {Lawrence}}]{alexandru2018finite}%
  \BibitemOpen
  \bibfield  {author} {\bibinfo {author} {\bibfnamefont {A.}~\bibnamefont
  {Alexandru}}, \bibinfo {author} {\bibfnamefont {G.}~\bibnamefont
  {Ba\ifmmode~\mbox{\c{s}}\else \c{s}\fi{}ar}}, \bibinfo {author}
  {\bibfnamefont {P.~F.}\ \bibnamefont {Bedaque}}, \bibinfo {author}
  {\bibfnamefont {H.}~\bibnamefont {Lamm}},\ and\ \bibinfo {author}
  {\bibfnamefont {S.}~\bibnamefont {Lawrence}},\ }\bibfield  {title} {\bibinfo
  {title} {Finite density {${\mathrm{QED}}_{1+1}$} near {L}efschetz thimbles},\
  }\href {https://doi.org/10.1103/PhysRevD.98.034506} {\bibfield  {journal}
  {\bibinfo  {journal} {Phys. Rev. D}\ }\textbf {\bibinfo {volume} {98}},\
  \bibinfo {pages} {034506} (\bibinfo {year} {2018}{\natexlab{b}})}\BibitemShut
  {NoStop}%
\bibitem [{\citenamefont {Mukherjee}\ and\ \citenamefont
  {Cristoforetti}(2014)}]{mukherjee2014lefschetz}%
  \BibitemOpen
  \bibfield  {author} {\bibinfo {author} {\bibfnamefont {A.}~\bibnamefont
  {Mukherjee}}\ and\ \bibinfo {author} {\bibfnamefont {M.}~\bibnamefont
  {Cristoforetti}},\ }\bibfield  {title} {\bibinfo {title} {Lefschetz thimble
  monte carlo for many-body theories: A hubbard model study},\ }\href
  {https://doi.org/10.1103/PhysRevB.90.035134} {\bibfield  {journal} {\bibinfo
  {journal} {Phys. Rev. B}\ }\textbf {\bibinfo {volume} {90}},\ \bibinfo
  {pages} {035134} (\bibinfo {year} {2014})}\BibitemShut {NoStop}%
\bibitem [{\citenamefont {Tanizaki}\ \emph {et~al.}(2016)\citenamefont
  {Tanizaki}, \citenamefont {Hidaka},\ and\ \citenamefont
  {Hayata}}]{tanizaki2016lefshetz}%
  \BibitemOpen
  \bibfield  {author} {\bibinfo {author} {\bibfnamefont {Y.}~\bibnamefont
  {Tanizaki}}, \bibinfo {author} {\bibfnamefont {Y.}~\bibnamefont {Hidaka}},\
  and\ \bibinfo {author} {\bibfnamefont {T.}~\bibnamefont {Hayata}},\
  }\bibfield  {title} {\bibinfo {title} {Lefschetz-thimble analysis of the sign
  problem in one-site fermion model},\ }\href
  {https://doi.org/10.1088/1367-2630/18/3/033002} {\bibfield  {journal}
  {\bibinfo  {journal} {New J. Phys.}\ }\textbf {\bibinfo {volume} {18}},\
  \bibinfo {pages} {033002} (\bibinfo {year} {2016})}\BibitemShut {NoStop}%
\bibitem [{\citenamefont {Ulybyshev}\ \emph {et~al.}(2020)\citenamefont
  {Ulybyshev}, \citenamefont {Winterowd},\ and\ \citenamefont
  {Zafeiropoulos}}]{ulybyshev2020lefschetz}%
  \BibitemOpen
  \bibfield  {author} {\bibinfo {author} {\bibfnamefont {M.}~\bibnamefont
  {Ulybyshev}}, \bibinfo {author} {\bibfnamefont {C.}~\bibnamefont
  {Winterowd}},\ and\ \bibinfo {author} {\bibfnamefont {S.}~\bibnamefont
  {Zafeiropoulos}},\ }\bibfield  {title} {\bibinfo {title} {Lefschetz thimbles
  decomposition for the hubbard model on the hexagonal lattice},\ }\href
  {https://doi.org/10.1103/PhysRevD.101.014508} {\bibfield  {journal} {\bibinfo
   {journal} {Phys. Rev. D}\ }\textbf {\bibinfo {volume} {101}},\ \bibinfo
  {pages} {014508} (\bibinfo {year} {2020})}\BibitemShut {NoStop}%
\bibitem [{\citenamefont {Stratonovich}(1958)}]{stratonovich1957method}%
  \BibitemOpen
  \bibfield  {author} {\bibinfo {author} {\bibfnamefont {R.}~\bibnamefont
  {Stratonovich}},\ }\bibfield  {title} {\bibinfo {title} {On a method of
  calculating quantum distribution functions},\ }\href@noop {} {\bibfield
  {journal} {\bibinfo  {journal} {Soviet Phys. Doklady}\ }\textbf {\bibinfo
  {volume} {2}},\ \bibinfo {pages} {416} (\bibinfo {year} {1958})}\BibitemShut
  {NoStop}%
\bibitem [{\citenamefont {Hubbard}(1959)}]{hubbard1959calculation}%
  \BibitemOpen
  \bibfield  {author} {\bibinfo {author} {\bibfnamefont {J.}~\bibnamefont
  {Hubbard}},\ }\bibfield  {title} {\bibinfo {title} {Calculation of partition
  functions},\ }\href {https://doi.org/10.1103/PhysRevLett.3.77} {\bibfield
  {journal} {\bibinfo  {journal} {Phys. Rev. Lett.}\ }\textbf {\bibinfo
  {volume} {3}},\ \bibinfo {pages} {77} (\bibinfo {year} {1959})}\BibitemShut
  {NoStop}%
\bibitem [{\citenamefont {Mishchenko}\ \emph {et~al.}(2021)\citenamefont
  {Mishchenko}, \citenamefont {Kato},\ and\ \citenamefont
  {Motome}}]{mishchenko_quantum_2021}%
  \BibitemOpen
  \bibfield  {author} {\bibinfo {author} {\bibfnamefont {P.~A.}\ \bibnamefont
  {Mishchenko}}, \bibinfo {author} {\bibfnamefont {Y.}~\bibnamefont {Kato}},\
  and\ \bibinfo {author} {\bibfnamefont {Y.}~\bibnamefont {Motome}},\
  }\bibfield  {title} {\bibinfo {title} {A quantum {Monte} {Carlo} method on
  asymptotic {L}efschetz thimbles for quantum spin systems: {An} application to
  the {Kitaev} model in a magnetic field},\ }\href
  {http://arxiv.org/abs/2106.07937} {\bibfield  {journal} {\bibinfo  {journal}
  {arXiv:2106.07937}\ } (\bibinfo {year} {2021})}\BibitemShut {NoStop}%
\bibitem [{\citenamefont {Shibata}\ and\ \citenamefont
  {Takagi}(1999)}]{shibata_note_1999}%
  \BibitemOpen
  \bibfield  {author} {\bibinfo {author} {\bibfnamefont {J.}~\bibnamefont
  {Shibata}}\ and\ \bibinfo {author} {\bibfnamefont {S.}~\bibnamefont
  {Takagi}},\ }\bibfield  {title} {\bibinfo {title} {A note on (spin-)
  coherent-state path integral},\ }\href
  {https://doi.org/10.1142/S0217979299000096} {\bibfield  {journal} {\bibinfo
  {journal} {Int. J. Mod. Phys. B}\ }\textbf {\bibinfo {volume} {13}},\
  \bibinfo {pages} {107} (\bibinfo {year} {1999})}\BibitemShut {NoStop}%
\bibitem [{\citenamefont {Auerbach}(1994)}]{auerbach}%
  \BibitemOpen
  \bibfield  {author} {\bibinfo {author} {\bibfnamefont {A.}~\bibnamefont
  {Auerbach}},\ }\href@noop {} {\emph {\bibinfo {title} {Interacting Electrons
  and Quantum Magnetism}}}\ (\bibinfo  {publisher} {Springer-Verlag},\ \bibinfo
  {year} {1994})\BibitemShut {NoStop}%
\bibitem [{\citenamefont {Wilson}\ and\ \citenamefont
  {Galitski}(2011)}]{galitski2011breakdown}%
  \BibitemOpen
  \bibfield  {author} {\bibinfo {author} {\bibfnamefont {J.~H.}\ \bibnamefont
  {Wilson}}\ and\ \bibinfo {author} {\bibfnamefont {V.}~\bibnamefont
  {Galitski}},\ }\bibfield  {title} {\bibinfo {title} {Breakdown of the
  coherent state path integral: Two simple examples},\ }\href
  {https://doi.org/10.1103/PhysRevLett.106.110401} {\bibfield  {journal}
  {\bibinfo  {journal} {Phys. Rev. Lett.}\ }\textbf {\bibinfo {volume} {106}},\
  \bibinfo {pages} {110401} (\bibinfo {year} {2011})}\BibitemShut {NoStop}%
\bibitem [{\citenamefont {Kordas}\ \emph {et~al.}(2016)\citenamefont {Kordas},
  \citenamefont {Kalantzis},\ and\ \citenamefont
  {Karanikas}}]{kordas2016coherent}%
  \BibitemOpen
  \bibfield  {author} {\bibinfo {author} {\bibfnamefont {G.}~\bibnamefont
  {Kordas}}, \bibinfo {author} {\bibfnamefont {D.}~\bibnamefont {Kalantzis}},\
  and\ \bibinfo {author} {\bibfnamefont {A.}~\bibnamefont {Karanikas}},\
  }\bibfield  {title} {\bibinfo {title} {Coherent-state path integrals in the
  continuum: The su(2) case},\ }\href
  {https://doi.org/https://doi.org/10.1016/j.aop.2016.05.012} {\bibfield
  {journal} {\bibinfo  {journal} {Annals of Physics}\ }\textbf {\bibinfo
  {volume} {372}},\ \bibinfo {pages} {226} (\bibinfo {year}
  {2016})}\BibitemShut {NoStop}%
\bibitem [{\citenamefont {Ran{\c{c}}on}(2020)}]{ranccon2020hubbard}%
  \BibitemOpen
  \bibfield  {author} {\bibinfo {author} {\bibfnamefont {A.}~\bibnamefont
  {Ran{\c{c}}on}},\ }\bibfield  {title} {\bibinfo {title}
  {Hubbard--stratonovich transformation and consistent ordering in the coherent
  state path integral: insights from stochastic calculus},\ }\href
  {https://doi.org/10.1088/1751-8121/ab6d3b} {\bibfield  {journal} {\bibinfo
  {journal} {J. Phys. A: Math. Theor.}\ }\textbf {\bibinfo {volume} {53}},\
  \bibinfo {pages} {105302} (\bibinfo {year} {2020})}\BibitemShut {NoStop}%
\bibitem [{\citenamefont {Di~Renzo}\ and\ \citenamefont
  {Eruzzi}(2015)}]{renzo2015thimble}%
  \BibitemOpen
  \bibfield  {author} {\bibinfo {author} {\bibfnamefont {F.}~\bibnamefont
  {Di~Renzo}}\ and\ \bibinfo {author} {\bibfnamefont {G.}~\bibnamefont
  {Eruzzi}},\ }\bibfield  {title} {\bibinfo {title} {Thimble regularization at
  work: From toy models to chiral random matrix theories},\ }\href
  {https://doi.org/10.1103/PhysRevD.92.085030} {\bibfield  {journal} {\bibinfo
  {journal} {Phys. Rev. D}\ }\textbf {\bibinfo {volume} {92}},\ \bibinfo
  {pages} {085030} (\bibinfo {year} {2015})}\BibitemShut {NoStop}%
\bibitem [{\citenamefont {Alexandru}\ \emph
  {et~al.}(2016{\natexlab{b}})\citenamefont {Alexandru}, \citenamefont {Basar},
  \citenamefont {Bedaque}, \citenamefont {Ridgway},\ and\ \citenamefont
  {Warrington}}]{alexandru2016sign}%
  \BibitemOpen
  \bibfield  {author} {\bibinfo {author} {\bibfnamefont {A.}~\bibnamefont
  {Alexandru}}, \bibinfo {author} {\bibfnamefont {G.}~\bibnamefont {Basar}},
  \bibinfo {author} {\bibfnamefont {P.~F.}\ \bibnamefont {Bedaque}}, \bibinfo
  {author} {\bibfnamefont {G.~W.}\ \bibnamefont {Ridgway}},\ and\ \bibinfo
  {author} {\bibfnamefont {N.~C.}\ \bibnamefont {Warrington}},\ }\bibfield
  {title} {\bibinfo {title} {Sign problem and monte carlo calculations beyond
  lefschetz thimbles},\ }\href {https://doi.org/10.1007/JHEP05(2016)053}
  {\bibfield  {journal} {\bibinfo  {journal} {J. High Energ. Phys.}\ }\textbf
  {\bibinfo {volume} {2016}}\bibinfo  {number} { (5)},\ \bibinfo {pages}
  {1}}\BibitemShut {NoStop}%
\bibitem [{\citenamefont {Fukuma}\ and\ \citenamefont
  {Umeda}(2017)}]{fukuma2017parallel}%
  \BibitemOpen
\bibfield  {number} {  }\bibfield  {author} {\bibinfo {author} {\bibfnamefont
  {M.}~\bibnamefont {Fukuma}}\ and\ \bibinfo {author} {\bibfnamefont
  {N.}~\bibnamefont {Umeda}},\ }\bibfield  {title} {\bibinfo {title} {Parallel
  tempering algorithm for integration over {L}efschetz thimbles},\ }\href
  {https://doi.org/10.1093/ptep/ptx081} {\bibfield  {journal} {\bibinfo
  {journal} {Prog. Theor. Exp. Phys.}\ }\textbf {\bibinfo {volume} {2017}},\
  \bibinfo {pages} {073B01} (\bibinfo {year} {2017})}\BibitemShut {NoStop}%
\bibitem [{\citenamefont {Alexandru}\ \emph
  {et~al.}(2017{\natexlab{b}})\citenamefont {Alexandru}, \citenamefont
  {Ba\ifmmode~\mbox{\c{s}}\else \c{s}\fi{}ar}, \citenamefont {Bedaque},\ and\
  \citenamefont {Ridgway}}]{alexandru2017schwinger}%
  \BibitemOpen
  \bibfield  {author} {\bibinfo {author} {\bibfnamefont {A.}~\bibnamefont
  {Alexandru}}, \bibinfo {author} {\bibfnamefont {G.}~\bibnamefont
  {Ba\ifmmode~\mbox{\c{s}}\else \c{s}\fi{}ar}}, \bibinfo {author}
  {\bibfnamefont {P.~F.}\ \bibnamefont {Bedaque}},\ and\ \bibinfo {author}
  {\bibfnamefont {G.}~\bibnamefont {Ridgway}},\ }\bibfield  {title} {\bibinfo
  {title} {Schwinger-{K}eldysh formalism on the lattice: A faster algorithm and
  its application to field theory},\ }\href
  {https://doi.org/10.1103/PhysRevD.95.114501} {\bibfield  {journal} {\bibinfo
  {journal} {Phys. Rev. D}\ }\textbf {\bibinfo {volume} {95}},\ \bibinfo
  {pages} {114501} (\bibinfo {year} {2017}{\natexlab{b}})}\BibitemShut
  {NoStop}%
\bibitem [{\citenamefont {Alexandru}\ \emph {et~al.}(2020)\citenamefont
  {Alexandru}, \citenamefont {Ba\ifmmode~\mbox{\c{s}}\else \c{s}\fi{}ar},
  \citenamefont {Bedaque},\ and\ \citenamefont
  {Warrington}}]{alexandru_complex_2020}%
  \BibitemOpen
  \bibfield  {author} {\bibinfo {author} {\bibfnamefont {A.}~\bibnamefont
  {Alexandru}}, \bibinfo {author} {\bibfnamefont {G.}~\bibnamefont
  {Ba\ifmmode~\mbox{\c{s}}\else \c{s}\fi{}ar}}, \bibinfo {author}
  {\bibfnamefont {P.~F.}\ \bibnamefont {Bedaque}},\ and\ \bibinfo {author}
  {\bibfnamefont {N.~C.}\ \bibnamefont {Warrington}},\ }\bibfield  {title}
  {\bibinfo {title} {Complex {Paths} {Around} {The} {Sign} {Problem}},\ }\href
  {http://arxiv.org/abs/2007.05436} {\bibfield  {journal} {\bibinfo  {journal}
  {arXiv:2007.05436}\ } (\bibinfo {year} {2020})}\BibitemShut {NoStop}%
\bibitem [{\citenamefont {Alexandru}\ \emph
  {et~al.}(2016{\natexlab{c}})\citenamefont {Alexandru}, \citenamefont
  {Ba\ifmmode~\mbox{\c{s}}\else \c{s}\fi{}ar}, \citenamefont {Bedaque},
  \citenamefont {Ridgway},\ and\ \citenamefont
  {Warrington}}]{Alexandru_estimator}%
  \BibitemOpen
  \bibfield  {author} {\bibinfo {author} {\bibfnamefont {A.}~\bibnamefont
  {Alexandru}}, \bibinfo {author} {\bibfnamefont {G.}~\bibnamefont
  {Ba\ifmmode~\mbox{\c{s}}\else \c{s}\fi{}ar}}, \bibinfo {author}
  {\bibfnamefont {P.~F.}\ \bibnamefont {Bedaque}}, \bibinfo {author}
  {\bibfnamefont {G.~W.}\ \bibnamefont {Ridgway}},\ and\ \bibinfo {author}
  {\bibfnamefont {N.~C.}\ \bibnamefont {Warrington}},\ }\bibfield  {title}
  {\bibinfo {title} {Fast estimator of {J}acobians in the {M}onte {C}arlo
  integration on {L}efschetz thimbles},\ }\href
  {https://link.aps.org/doi/10.1103/PhysRevD.93.094514} {\bibfield  {journal}
  {\bibinfo  {journal} {Phys. Rev. D}\ }\textbf {\bibinfo {volume} {93}},\
  \bibinfo {pages} {094514} (\bibinfo {year} {2016}{\natexlab{c}})}\BibitemShut
  {NoStop}%
\bibitem [{\citenamefont {Duane}\ \emph {et~al.}(1987)\citenamefont {Duane},
  \citenamefont {Kennedy}, \citenamefont {Pendleton},\ and\ \citenamefont
  {Roweth}}]{DUANE1987216}%
  \BibitemOpen
  \bibfield  {author} {\bibinfo {author} {\bibfnamefont {S.}~\bibnamefont
  {Duane}}, \bibinfo {author} {\bibfnamefont {A.}~\bibnamefont {Kennedy}},
  \bibinfo {author} {\bibfnamefont {B.~J.}\ \bibnamefont {Pendleton}},\ and\
  \bibinfo {author} {\bibfnamefont {D.}~\bibnamefont {Roweth}},\ }\bibfield
  {title} {\bibinfo {title} {Hybrid monte carlo},\ }\href
  {https://doi.org/https://doi.org/10.1016/0370-2693(87)91197-X} {\bibfield
  {journal} {\bibinfo  {journal} {Physics Letters B}\ }\textbf {\bibinfo
  {volume} {195}},\ \bibinfo {pages} {216} (\bibinfo {year}
  {1987})}\BibitemShut {NoStop}%
\bibitem [{\citenamefont {Radcliffe}(1971)}]{radcliffe_properties_1971}%
  \BibitemOpen
  \bibfield  {author} {\bibinfo {author} {\bibfnamefont {J.~M.}\ \bibnamefont
  {Radcliffe}},\ }\bibfield  {title} {\bibinfo {title} {Some properties of
  coherent spin states},\ }\href {https://doi.org/10.1088/0305-4470/4/3/009}
  {\bibfield  {journal} {\bibinfo  {journal} {J. Phys. A: Gen. Phys.}\ }\textbf
  {\bibinfo {volume} {4}},\ \bibinfo {pages} {313} (\bibinfo {year}
  {1971})}\BibitemShut {NoStop}%
\bibitem [{\citenamefont {Kochetov}(1995)}]{kochetov19952}%
  \BibitemOpen
  \bibfield  {author} {\bibinfo {author} {\bibfnamefont {E.}~\bibnamefont
  {Kochetov}},\ }\bibfield  {title} {\bibinfo {title} {{SU}(2) coherent-state
  path integral},\ }\href {https://doi.org/10.1063/1.530913} {\bibfield
  {journal} {\bibinfo  {journal} {J. Math. Phys.}\ }\textbf {\bibinfo {volume}
  {36}},\ \bibinfo {pages} {4667} (\bibinfo {year} {1995})}\BibitemShut
  {NoStop}%
\bibitem [{\citenamefont {Takano}(1985)}]{takano1985monte}%
  \BibitemOpen
  \bibfield  {author} {\bibinfo {author} {\bibfnamefont {H.}~\bibnamefont
  {Takano}},\ }\bibfield  {title} {\bibinfo {title} {Monte {C}arlo method for
  quantum spin systems based on the {B}loch coherent state representation},\
  }\href {https://doi.org/10.1143/PTP.73.332} {\bibfield  {journal} {\bibinfo
  {journal} {Prog. Theor. Phys.}\ }\textbf {\bibinfo {volume} {73}},\ \bibinfo
  {pages} {332} (\bibinfo {year} {1985})}\BibitemShut {NoStop}%
\bibitem [{\citenamefont {Garg}\ \emph {et~al.}(2001)\citenamefont {Garg},
  \citenamefont {Kochetov}, \citenamefont {Park},\ and\ \citenamefont
  {Stone}}]{garg_spin_2001}%
  \BibitemOpen
  \bibfield  {author} {\bibinfo {author} {\bibfnamefont {A.}~\bibnamefont
  {Garg}}, \bibinfo {author} {\bibfnamefont {E.}~\bibnamefont {Kochetov}},
  \bibinfo {author} {\bibfnamefont {K.-S.}\ \bibnamefont {Park}},\ and\
  \bibinfo {author} {\bibfnamefont {M.}~\bibnamefont {Stone}},\ }\bibfield
  {title} {\bibinfo {title} {Spin coherent-state path integrals and the
  instanton calculus},\ }\href {http://arxiv.org/abs/cond-mat/0111139}
  {\bibfield  {journal} {\bibinfo  {journal} {arXiv:cond-mat/0111139}\ }
  (\bibinfo {year} {2001})}\BibitemShut {NoStop}%
\bibitem [{\citenamefont {Tanizaki}\ \emph {et~al.}(2017)\citenamefont
  {Tanizaki}, \citenamefont {Nishimura},\ and\ \citenamefont
  {J.~M.~Verbaarschot}}]{tanizaki_gradient_2017}%
  \BibitemOpen
  \bibfield  {author} {\bibinfo {author} {\bibfnamefont {Y.}~\bibnamefont
  {Tanizaki}}, \bibinfo {author} {\bibfnamefont {H.}~\bibnamefont
  {Nishimura}},\ and\ \bibinfo {author} {\bibfnamefont {J.}~\bibnamefont
  {J.~M.~Verbaarschot}},\ }\bibfield  {title} {\bibinfo {title} {Gradient flows
  without blow-up for {L}efschetz thimbles},\ }\href
  {https://doi.org/10.1007/JHEP10(2017)100} {\bibfield  {journal} {\bibinfo
  {journal} {J. High Energy Phys{.}}\ }\textbf {\bibinfo {volume} {2017}},\
  \bibinfo {pages} {100} (\bibinfo {year} {2017})}\BibitemShut {NoStop}%
\bibitem [{cod()}]{code}%
  \BibitemOpen
  \href@noop {} {}\bibinfo {note} {The code developed for this work is
  available upon request.}\BibitemShut {Stop}%
\end{thebibliography}%


%

\newpage
\onecolumngrid

\begin{appendix}
\section{Derivation of Spin Coherent State Path Integral} \label{section:appA}
In this appendix, we derive the spin coherent state path integral.
\subsection{The Partition Function}
To derive the spin coherent state path integral, we begin with the definition of the partition function:
\begin{equation*}
    \mathcal{Z} = \mathrm{Tr}\{e^{-\beta \hat{H}}\}.
\end{equation*}
We restate the spin coherent state resolution of the identity here for convenience
\begin{equation} \label{eq:res-identity}
    I = \frac{2S+1}{\pi}\int_{\mathbb{R}^2} \frac{\mathrm{d}^2\mu}{(1+\vert\mu\vert^2)^2}\ket{\mu}\bra{\mu}.
\end{equation}
Now, as $\hat{H}$ commutes with itself, we can insert the spin coherent state resolution of the identity from Eq.~(\ref{eq:res-identity}) between each imaginary-time step and rewrite this expression as
\begin{align}
    \mathcal{Z} &= \mathrm{Tr}\left\{\left(e^{-(\beta/T) \hat{H}}\right) ^T\right\}\nonumber\\
    &= \mathrm{Tr}\left\{\left(e^{-(\beta/T) \hat{H}}\left(\frac{2S+1}{\pi}\int_{\mathbb{R}^2} \frac{\mathrm{d}^2\mu}{\left(1+\vert \mu\vert^2\right)^2}\ket{\mu}\bra{\mu}\right)\right) ^T\right\}\nonumber\\
    &=\left(\prod_{j=0}^{T-1}\frac{2S+1}{\pi}\int_{\mathbb{R}^2} \frac{\mathrm{d}^2\mu_j}{\left(1+\vert \mu_j\vert^2\right)^2}\right)\left(\prod_{j=0}^{T-1}\bra{\mu_{j+1}}e^{-\frac{\beta}{T}\hat{H}}\ket{\mu_j}\right) \label{eq:partitionpathint}
\end{align}
where, enforcing the periodicity imposed by the trace, we have $\mu_T=\mu_0.$ Next we expand the exponential out to first order in $\beta/T$ and get that 
\begin{align}
    \bra{\mu_{j+1}}e^{-\frac{\beta}{T}\hat{H}}\ket{\mu_j}
    &= \bra{\mu_{j+1}}\left(I-\frac{\beta}{T} \hat{H} + \mathcal{O}\left(\left(\frac{\beta}{T}\right)^2\right)\right)\ket{\mu_j} \nonumber\\
    &= \langle \mu_{j+1}\vert \mu_j\rangle \Bigg(1 - \frac{\beta}{T} H^{cl}(\overline{\mu}_{j+1}, \mu_j)+\mathcal{O}\left(\left(\beta/T\right)^2\right)\Bigg),
\end{align}
where $H^{cl}(\overline{\mu}_{j+1}, \mu_j)\equiv \bra{\mu_{j+1}}H\ket{\mu_j}/\langle \mu_{j+1}\vert \mu_j\rangle$ will be examined in detail in the next section.  Now, using that the overlap of two spin coherent states $\ket{\mu}$ and $\ket{\mu'}$ is $\langle \mu'\vert \mu\rangle = (1+\overline{\mu}'\mu)^{2S}/((1+\vert \mu\vert^2)(1+\vert \mu'\vert^2))^S$ \cite{radcliffe_properties_1971}, we can rewrite this expression as
\begin{align}
    \frac{(1+\overline{\mu}_{j+1}\mu_j)^{2S}}{((1+\vert \mu_{j+1}\vert^2)(1+\vert \mu_j\vert^2))^S} \Bigg(1 - (\beta/T) H^{cl}(\overline{\mu}_{j+1}, \mu_j)+\mathcal{O}\left(\left(\beta/T\right)^2\right)\Bigg)
\end{align}

The next task is to simplify the inner product in front. To make this possible, we make the standard assumption in deriving path integral expansions that $\vert \mu_{j+1}-\mu_j \vert\equiv \vert \delta_{j+1} \vert= \mathcal{O}(\beta/T).$ There is analytical evidence that this assumption is not mathematically well-founded for spin-coherent state path integrals \cite{auerbach, shibata_note_1999,garg_spin_2001}, but we, like much of the literature making use of this technique, shall proceed despite that.
Thus, making that assumption, we can write that
\begin{align}
    \frac{(1+\overline{\mu}_{j+1}\mu_j)^{2S}}{((1+\vert \mu_{j+1}\vert^2)(1+\vert \mu_j\vert^2))^S}
    &= 1- \frac{((1+\vert \mu_{j+1}\vert^2)(1+\vert \mu_j\vert^2))^S-(1+\overline{\mu}_{j+1}\mu_j)^{2S}}{((1+\vert \mu_{j+1}\vert^2)(1+\vert \mu_j\vert^2))^S} \nonumber\\
    &= 1 + S\frac{ \left(\overline{\delta}_{j+1}\mu_{j}-\delta_{j+1}\overline{\mu}_j\right)(1+\vert \mu_j\vert^2)^{2S-1}}{((1+\vert \mu_{j+1}\vert^2)(1+\vert \mu_j\vert^2))^S}+\mathcal{O}((\beta/T)^2)\nonumber\\
    &= 1 + S\frac{ \left(\overline{\delta}_{j+1}\mu_{j}-\delta_{j+1}\overline{\mu}_j\right)}{(1+\vert \mu_j\vert^2)}+\mathcal{O}((\beta/T)^2),
\end{align}
where to get between the second and third, as well as third and fourth lines we use the binomial expansion.

Multiplying these two together, we finally get that

\begin{align}
    \bra{\mu_{j+1}}e^{-\beta H/T}\ket{\mu_j} &= 1 + S\frac{ \left(\overline{\delta}_{j+1}\mu_{j}-\delta_{j+1}\overline{\mu}_j\right)}{(1+\vert \mu_j\vert^2)}  - \frac{\beta}{T}H^{cl}(\overline{\mu}_{j+1},\mu_j)+\mathcal{O}\left(\left(\frac{\beta}{T}\right)^2\right)\nonumber\\
    &= \exp\Bigg\{S\frac{ \left(\overline{\delta}_{j+1}\mu_{j}-\delta_{j+1}\overline{\mu}_j\right)}{(1+\vert \mu_j\vert^2)}  - \frac{\beta}{T}H^{cl}(\overline{\mu}_{j},\mu_j)\Bigg\}+\mathcal{O}\left(\left(\frac{\beta}{T}\right)^2\right).
\end{align}

Plugging this back into Eq.~(\ref{eq:partitionpathint}), we get that 
\begin{align}
    \mathcal{Z} &= \left(\prod_{j=0}^{T-1}\frac{2S+1}{\pi}\int_{\mathbb{R}^2} \frac{\mathrm{d}^2\mu_j}{\left(1+\vert \mu_j\vert^2\right)^2}\right)e^{-\mathcal{S}[\{\mu_j\}]}
   + \mathcal{O}(\beta/T),
\end{align}
where 
\begin{equation}\label{eq:SCS_discreteaction}
\mathcal{S}[\{\mu_j\}] \equiv \sum_{j=0}^{T-1} \left(-S\frac{ \left(\overline{\delta}_{j+1}\mu_{j}-\delta_{j+1}\overline{\mu}_j\right)}{(1+\vert \mu_j\vert^2)}+\frac{\beta}{T}H^{cl}(\overline{\mu}_j,\mu_j)\right).
\end{equation}
Letting $x\equiv \mathrm{Re}\mu$ and $y\equiv \mathrm{Im}\mu,$ we arrive at
\begin{align}
        \mathcal{Z} = \Bigg(\prod_{j=0}^{T-1}\frac{2S+1}{\pi}\int_{\mathbb{R}^2} \frac{\mathrm{d}x_j\mathrm{d}y_j}{\left(1+ x_j^2+y_j^2\right)^2}&\Bigg)e^{-\mathcal{S}[\{x_j\},\{y_j\}]}+ \mathcal{O}(\beta/T),
\end{align}
\begin{align}
\mathcal{S}[\{x_j\},\{y_j\}]\equiv 
&\sum_{j=0}^{T-1} \bigg(2iS\frac{ \left((y_{j+1}-y_j)x_j-(x_{j+1}-x_j)y_j\right)}{(1+x_j^2+y_j^2)}+\frac{\beta}{T}H^{cl}(\overline{\mu}_j,\mu_j)\bigg) \nonumber\\
&= \sum_{j=0}^{T-1} \bigg(2iS\frac{ \left(y_{j+1}x_j-x_{j+1}y_j\right)}{(1+x_j^2+y_j^2)}+\frac{\beta}{T}H^{cl}(\overline{\mu}_j,\mu_j)\bigg)
\end{align}

This action is similar to that of \cite{kochetov19952}, but, importantly for our purposes, has no singularities over $\mathbb{R}^{2T}.$ Only when this action is continued to $\mathbb{C}^{2T}$ will singularities emerge.

To get the continuum limit, which we do not actually use in our analysis, but is nevertheless the most compact and aesthetically pleasing way of presenting the path integral, we then take $T\rightarrow \infty$ and notice that as $\delta_j = \mathcal{O}(\beta/T)$ by assumption, we can define derivatives with respect to that parameter. Thus, we get that 
\begin{equation}
    \mathcal{Z} = \int \mathcal{D} x\mathcal{D}y e^{-\mathcal{S}[x,y]},
\end{equation}
where 
\begin{equation}
    \mathcal{S}[x, y] =\int_0^\tau \mathrm{d}\tau \left(2iS \frac{\dot{y}x-\dot{x}y}{1+x^2+y^2}+H^{\mathrm{cl}}(x,y)\right), 
\end{equation}
and $H^{\mathrm{cl}}(x,y)\equiv H^{\mathrm{cl}}(x-iy, x+iy).$

\subsection{Classical Hamiltonian Elements}
One thing that we have not done yet is calculate $H^{cl}(x,y)$ explicitly. To do this, notice that $\ket{\uparrow} = (\ket{+1/2})^{\otimes 2S}$ and that \begin{align}
\label{eq:lowering_operator}
\hat{S}_- = (\hat{\sigma}_x - i\hat{\sigma}_y) \otimes I^{\otimes(2S-1)} + I \otimes (\hat{\sigma}_x - i\hat{\sigma}_y) \otimes I^{\otimes(2S-2)}+\cdots +I^{\otimes 2S-1}\otimes (\hat{\sigma}_x - i\hat{\sigma}_y).
\end{align} 
The definition of a spin coherent state is 
\begin{equation}
    \label{eq:exp_lowering_operator}
    \ket{\mu} = \frac{1}{(1+\vert\mu\vert^2)^S}\exp \{\mu\hat{S}_-\} \ket{\uparrow}.
\end{equation}
Plugging Eq.~(\ref{eq:lowering_operator}) and the definition of $\ket{\uparrow}$ into Eq.~(\ref{eq:exp_lowering_operator}), we can utilize the commutivity of the terms in Eq.~(\ref{eq:lowering_operator}) to get that
\begin{align*}
    \ket{\mu} &= \frac{1}{(1+\vert\mu\vert^2)^S} \bigotimes_1^{2S}\left(\exp\{\mu (\hat{\sigma}_x-i\hat{\sigma}_y)\} \ket{+1/2}\right).
\end{align*}
We can write out the state in the spin-1/2 space explicitly, using that $(\hat{\sigma}_x-i\hat{\sigma}_y)^2=0,$ to obtain that 
\begin{align*}
    \left(\exp\{\mu (\hat{\sigma}_x-i\hat{\sigma}_y)\} \ket{+1/2}\right) &= \ket{+1/2} + \mu \ket{-1/2}.
\end{align*}
Now, let $\mu \equiv x+iy,$ and notice that we can write $\hat{S}_x,$ $\hat{S}_y,$ and $\hat{S}_\mu$ in the same way as $\hat{S}_-.$ Thus, We can use this expression to finally get the classical Hamiltonians corresponding to the spin operators: 
\begin{align}
    \hat{S}_x: \ H^{\mathrm{cl}}(x,y) &= 2S\frac{x}{1+x^2+y^2}\\
    \hat{S}_y: \ H^{\mathrm{cl}}(x,y) &= 2S\frac{y}{1+x^2+y^2}\\
    \hat{S}_z: \ H^{\mathrm{cl}}(x,y) &= S\frac{1-x^2-y^2}{1+x^2+y^2}.
\end{align}
\subsection{Multi-particle Systems}
Adapting the above derivations to systems with multiple particles is straightforward. Doing so, we get that for an $n$-particle system the path integral is:
\begin{equation}
    \mathcal{Z} = \int \left(\prod_{j=1}^n\mathcal{D}x^{(j)}\mathcal{D}y^{(j)}\right)e^{-\mathcal{S}\left[\{x^{(j)}\},\{y^{(j)}\}\right]}
\end{equation}
where 
\begin{align}
    \mathcal{S}\left[\{x^{(j)}\},\{y^{(j)}\}\right] =
    \int_0^\tau \mathrm{d}\tau \Bigg(2iS \sum_{j=1}^{n}\frac{\dot{y}^{(j)}x^{(j)}-\dot{x}^{(j)}y^{(j)}}{1+(x^{(j)})^2+(y^{(j)})^2}
    H^{\mathrm{cl}}\left(\{x^{(j)}\},\{y^{(j)}\}\right)\Bigg)
\end{align}
and $H^{\mathrm{cl}}(\{x^{(j)}\},\{y^{(j)}\})$ is obtained by replacing $\hat{S}_a$ acting on the $j^{\mathrm{th}}$ particle by its corresponding single-particle classical Hamiltonian $H^{cl}(x^{(j)},y^{(j)}),$ i.e. 
\begin{align}
    \hat{S}_z \otimes \hat{S}_x \rightarrow  \left(S\frac{1-(x^{(1)})^2-(y^{(1)})^2}{1+(x^{(1)})^2+(y^{(1)})^2}\right)\left(\frac{2Sx^{(2)}}{1+(x^{(2)})^2+(y^{(2)})^2}\right).
\end{align}

\end{appendix}

\end{document}


\renewcommand{\citenumfont}[1]{S#1} 
\renewcommand{\bibnumfmt}[1]{[S#1]} 

\title{Supplemental Material for `Lefschetz Thimble Quantum Monte Carlo for Spin Systems'}

\maketitle 

In this Supplemental Material, we derive the spin coherent state path integral (Sec.~\ref{section:appA}) and specify the details of our numerical implementation of standard path integral quantum Monte Carlo for the example of the frustrated spin triplet (Sec.~\ref{section:appB})

\hypersetup{linkcolor=black}
\tableofcontents

\section{Derivation of Spin Coherent State Path Integral} \label{section:appA}
\subsection{The Partition Function}
To derive the spin coherent state path integral, we begin with the definition of the partition function:
\begin{equation*}
    \mathcal{Z} = \mathrm{Tr}\{e^{-\beta \hat{H}}\}.
\end{equation*}
We restate the spin coherent state resolution of the identity here for convenience
\begin{equation} \label{eq:res-identity}
    I = \frac{2S+1}{\pi}\int_{\mathbb{R}^2} \frac{\mathrm{d}^2\mu}{(1+\vert\mu\vert^2)^2}\ket{\mu}\bra{\mu}.
\end{equation}
Now, as $\hat{H}$ commutes with itself, we can insert the spin coherent state resolution of the identity from Eq.~(\ref{eq:res-identity}) between each imaginary-time step and rewrite this expression as
\begin{align}
    \mathcal{Z} &= \mathrm{Tr}\left\{\left(e^{-(\beta/T) \hat{H}}\right) ^T\right\}\nonumber\\
    &= \mathrm{Tr}\left\{\left(e^{-(\beta/T) \hat{H}}\left(\frac{2S+1}{\pi}\int_{\mathbb{R}^2} \frac{\mathrm{d}^2\mu}{\left(1+\vert \mu\vert^2\right)^2}\ket{\mu}\bra{\mu}\right)\right) ^T\right\}\nonumber\\
    &=\left(\prod_{j=0}^{T-1}\frac{2S+1}{\pi}\int_{\mathbb{R}^2} \frac{\mathrm{d}^2\mu_j}{\left(1+\vert \mu_j\vert^2\right)^2}\right)\left(\prod_{j=0}^{T-1}\bra{\mu_{j+1}}e^{-\frac{\beta}{T}\hat{H}}\ket{\mu_j}\right) \label{eq:partitionpathint}
\end{align}
where, enforcing the periodicity imposed by the trace, we have $\mu_T=\mu_0.$ Next we expand the exponential out to first order in $\beta/T$ and get that 
\begin{align*}
    \bra{\mu_{j+1}}e^{-\frac{\beta}{T}\hat{H}}\ket{\mu_j} &= \bra{\mu_{j+1}}\left(I-\frac{\beta}{T} \hat{H} + \mathcal{O}\left(\left(\frac{\beta}{T}\right)^2\right)\right)\ket{\mu_j}\\
    &= \langle \mu_{j+1}\vert \mu_j\rangle \Bigg(1 - \frac{\beta}{T} H^{cl}(\overline{\mu}_{j+1}, \mu_j)+\mathcal{O}\left(\left(\beta/T\right)^2\right)\Bigg),
\end{align*}
where $H^{cl}(\overline{\mu}_{j+1}, \mu_j)\equiv \bra{\mu_{j+1}}H\ket{\mu_j}/\langle \mu_{j+1}\vert \mu_j\rangle$ will be examined in detail in the next section.  Now, using that the overlap of two spin coherent states $\ket{\mu}$ and $\ket{\mu'}$ is $\langle \mu'\vert \mu\rangle = (1+\overline{\mu}'\mu)^{2S}/((1+\vert \mu\vert^2)(1+\vert \mu'\vert^2))^S$ \cite{radcliffe_properties_1971}, we can rewrite this expression as
\begin{align*}
    \frac{(1+\overline{\mu}_{j+1}\mu_j)^{2S}}{((1+\vert \mu_{j+1}\vert^2)(1+\vert \mu_j\vert^2))^S} &\Bigg(1 - (\beta/T) H^{cl}(\overline{\mu}_{j+1}, \mu_j)+\mathcal{O}\left(\left(\beta/T\right)^2\right)\Bigg)
\end{align*}

The next task is to simplify the inner product in front. To make this possible, we make the standard assumption in deriving path integral expansions that $\vert \mu_{j+1}-\mu_j \vert\equiv \vert \delta_{j+1} \vert= \mathcal{O}(\beta/T).$ There is analytical evidence that this assumption is not mathematically well-founded for spin-coherent state path integrals \cite{auerbach, shibata_note_1999,garg_spin_2001}, but we, like much of the literature making use of this technique, shall proceed despite that.
Thus, making that assumption, we can write that
\begin{align*}
    \frac{(1+\overline{\mu}_{j+1}\mu_j)^{2S}}{((1+\vert \mu_{j+1}\vert^2)(1+\vert \mu_j\vert^2))^S} &= 1- \frac{((1+\vert \mu_{j+1}\vert^2)(1+\vert \mu_j\vert^2))^S-(1+\overline{\mu}_{j+1}\mu_j)^{2S}}{((1+\vert \mu_{j+1}\vert^2)(1+\vert \mu_j\vert^2))^S} \\
    &= 1 + S\frac{ \left(\overline{\delta}_{j+1}\mu_{j}-\delta_{j+1}\overline{\mu}_j\right)(1+\vert \mu_j\vert^2)^{2S-1}}{((1+\vert \mu_{j+1}\vert^2)(1+\vert \mu_j\vert^2))^S}+\mathcal{O}((\beta/T)^2)\\
    &= 1 + S\frac{ \left(\overline{\delta}_{j+1}\mu_{j}-\delta_{j+1}\overline{\mu}_j\right)}{(1+\vert \mu_j\vert^2)}+\mathcal{O}((\beta/T)^2),
\end{align*}
where to get between the first and second, as well as second and third lines we use the binomial expansion.

Multiplying these two together, we finally get that
\begin{align}
    \bra{\mu_{j+1}}e^{-\beta H/T}\ket{\mu_j} &= 1 + S\frac{ \left(\overline{\delta}_{j+1}\mu_{j}-\delta_{j+1}\overline{\mu}_j\right)}{(1+\vert \mu_j\vert^2)}  - \frac{\beta}{T}H^{cl}(\overline{\mu}_{j+1},\mu_j)+\mathcal{O}\left(\left(\frac{\beta}{T}\right)^2\right)\nonumber\\
    &= \exp\Bigg\{S\frac{ \left(\overline{\delta}_{j+1}\mu_{j}-\delta_{j+1}\overline{\mu}_j\right)}{(1+\vert \mu_j\vert^2)}  - \frac{\beta}{T}H^{cl}(\overline{\mu}_{j},\mu_j)\Bigg\}+\mathcal{O}\left(\left(\frac{\beta}{T}\right)^2\right).
\end{align}
Plugging this back into Eq.~(\ref{eq:partitionpathint}), we get that 
\begin{align}
    \mathcal{Z} &= \left(\prod_{j=0}^{T-1}\frac{2S+1}{\pi}\int_{\mathbb{R}^2} \frac{\mathrm{d}^2\mu_j}{\left(1+\vert \mu_j\vert^2\right)^2}\right)e^{-\mathcal{S}[\{\mu_j\}]}
   + \mathcal{O}(\beta/T),
\end{align}
where 
\begin{equation}\label{eq:SCS_discreteaction}
\mathcal{S}[\{\mu_j\}] \equiv \sum_{j=0}^{T-1} \left(-S\frac{ \left(\overline{\delta}_{j+1}\mu_{j}-\delta_{j+1}\overline{\mu}_j\right)}{(1+\vert \mu_j\vert^2)}+\frac{\beta}{T}H^{cl}(\overline{\mu}_j,\mu_j)\right).
\end{equation}
Letting $x\equiv \mathrm{Re}\mu$ and $y\equiv \mathrm{Im}\mu,$ we arrive at
\begin{equation}
        \mathcal{Z} = \left(\prod_{j=0}^{T-1}\frac{2S+1}{\pi}\int_{\mathbb{R}^2} \frac{\mathrm{d}x_j\mathrm{d}y_j}{\left(1+ x_j^2+y_j^2\right)^2}\right)e^{-\mathcal{S}[\{x_j\},\{y_j\}]}
   + \mathcal{O}(\beta/T),
\end{equation}
\begin{equation}
\mathcal{S}[\{x_j\},\{y_j\}] \equiv \sum_{j=0}^{T-1} \left(2iS\frac{ \left((y_{j+1}-y_j)x_j-(x_{j+1}-x_j)y_j\right)}{(1+x_j^2+y_j^2)}+\frac{\beta}{T}H^{cl}(\overline{\mu}_j,\mu_j)\right).
\end{equation}

This action is similar to that of \cite{kochetov19952}, but, importantly for our purposes, has no singularities over $\mathbb{R}^{2T}.$ Only when this action is continued to $\mathbb{C}^{2T}$ will singularities emerge.

To get the continuum limit, which we do not actually use in our analysis, but is nevertheless the most compact and aesthetically pleasing way of presenting the path integral, we then take $T\rightarrow \infty$ and notice that as $\delta_j = \mathcal{O}(\beta/T)$ by assumption, we can define derivatives with respect to that parameter. Thus, we get that 
\begin{equation}
    \mathcal{Z} = \int \mathcal{D} x\mathcal{D}y e^{-\mathcal{S}[x,y]},
\end{equation}
where 
\begin{equation}
    \mathcal{S}[x, y] =\int_0^\tau \mathrm{d}\tau \left(2iS \frac{\dot{y}x-\dot{x}y}{1+x^2+y^2}+H^{\mathrm{cl}}(x,y)\right), 
\end{equation}
and $H^{\mathrm{cl}}(x,y)\equiv H^{\mathrm{cl}}(x-iy, x+iy).$

\subsection{Classical Hamiltonian Elements}
One thing that we have not done yet is calculate $H^{cl}(x,y)$ explicitly. To do this, notice that $\ket{\uparrow} = (\ket{+1/2})^{\otimes 2S}$ and that \begin{align}
\label{eq:lowering_operator}
\hat{S}_- &= (\hat{\sigma}_x - i\hat{\sigma}_y) \otimes I^{\otimes(2S-1)} + I \otimes (\hat{\sigma}_x - i\hat{\sigma}_y) \otimes I^{\otimes(2S-2)}\nonumber\\&\qquad+\cdots +I^{\otimes 2S-1}\otimes (\hat{\sigma}_x - i\hat{\sigma}_y).
\end{align} 
The definition of a spin coherent state is 
\begin{equation}
    \label{eq:exp_lowering_operator}
    \ket{\mu} = \frac{1}{(1+\vert\mu\vert^2)^S}\exp \{\mu\hat{S}_-\} \ket{\uparrow}.
\end{equation}
Plugging Eq.~(\ref{eq:lowering_operator}) and the definition of $\ket{\uparrow}$ into Eq.~(\ref{eq:exp_lowering_operator}), we can utilize the commutivity of the terms in Eq.~(\ref{eq:lowering_operator}) to get that
\begin{align*}
    \ket{\mu} &= \frac{1}{(1+\vert\mu\vert^2)^S} \bigotimes_1^{2S}\left(\exp\{\mu (\hat{\sigma}_x-i\hat{\sigma}_y)\} \ket{+1/2}\right).
\end{align*}
We can write out the state in the spin-1/2 space explicitly, using that $(\hat{\sigma}_x-i\hat{\sigma}_y)^2=0,$ to obtain that 
\begin{align*}
    \left(\exp\{\mu (\hat{\sigma}_x-i\hat{\sigma}_y)\} \ket{+1/2}\right) &= \ket{+1/2} + \mu \ket{-1/2}.
\end{align*}
Now, let $\mu \equiv x+iy,$ and notice that we can write $\hat{S}_x,$ $\hat{S}_y,$ and $\hat{S}_\mu$ in the same way as $\hat{S}_-.$ Thus, We can use this expression to finally get the classical Hamiltonians corresponding to the spin operators: 
\begin{align}
    \hat{S}_x: \ H^{\mathrm{cl}}(x,y) &= 2S\frac{x}{1+x^2+y^2}\\
    \hat{S}_y: \ H^{\mathrm{cl}}(x,y) &= 2S\frac{y}{1+x^2+y^2}\\
    \hat{S}_z: \ H^{\mathrm{cl}}(x,y) &= S\frac{1-x^2-y^2}{1+x^2+y^2}.
\end{align}
\subsection{Multi-particle Systems}
Adapting the above derivations to systems with multiple particles is straightforward. Doing so, we get that for an $n$-particle system the path integral is:
\begin{equation}
    \mathcal{Z} = \int \left(\prod_{j=1}^n\mathcal{D}x^{(j)}\mathcal{D}y^{(j)}\right)e^{-\mathcal{S}\left[\{x^{(j)}\},\{y^{(j)}\}\right]}
\end{equation}
where 
\begin{align}
    \mathcal{S}\left[\{x^{(j)}\},\{y^{(j)}\}\right] &= \int_0^\tau \mathrm{d}\tau \Bigg(2iS \sum_{j=1}^{n}\frac{\dot{y}^{(j)}x^{(j)}-\dot{x}^{(j)}y^{(j)}}{1+(x^{(j)})^2+(y^{(j)})^2}\nonumber+H^{\mathrm{cl}}\left(\{x^{(j)}\},\{y^{(j)}\}\right)\Bigg)
\end{align}
and $H^{\mathrm{cl}}(\{x^{(j)}\},\{y^{(j)}\})$ is obtained by replacing $\hat{S}_a$ acting on the $j^{\mathrm{th}}$ particle by its corresponding single-particle classical Hamiltonian $H^{cl}(x^{(j)},y^{(j)}),$ i.e. 
\begin{align*}
    \hat{S}_z \otimes \hat{S}_x \rightarrow \left(S\frac{1-(x^{(1)})^2-(y^{(1)})^2}{1+(x^{(1)})^2+(y^{(1)})^2}\right)\left(\frac{2Sx^{(2)}}{1+(x^{(2)})^2+(y^{(2)})^2}\right).
\end{align*}

\section{Path Integral Quantum Monte Carlo Implementation}\label{section:appB}
To do path integral Quantum Monte Carlo on a frustrated triplet with Hamiltonian 
\begin{equation}
    \hat{H} = \hat{S}_{1,z}\hat{S}_{2,z}+\hat{S}_{2,z}\hat{S}_{3,z}+\hat{S}_{1,z}\hat{S}_{3,z} + \hat{S}_{1,x}\hat{S}_{2,x} + \hat{S}_{2,x}\hat{S}_{3,x} + \hat{S}_{1,x}\hat{S}_{3,x},
\end{equation}
we decompose $\hat{H}$ into a diagonal and off-diagonal part with $\hat{H}_d \equiv \hat{S}_{1,z}\hat{S}_{2,z}+\hat{S}_{2,z}\hat{S}_{3,z}+\hat{S}_{1,z}\hat{S}_{3,z}$ and $\hat{H}_o \equiv  \hat{S}_{1,x}\hat{S}_{2,x} + \hat{S}_{2,x}\hat{S}_{3,x} + \hat{S}_{1,x}\hat{S}_{3,x}$ respectively. We then use the approximation 
\begin{align}
    e^{-\beta(\hat{H}_d + \hat{H}_o)} &= \left(e^{-\beta(\hat{H}_d + \hat{H}_o)/T}\right)^T \nonumber\\
    &\approx \left(e^{-\beta \hat{H}_d/T}e^{-\beta \hat{H}_o/T}\right)^T,
\end{align}
for large $T$.

Now, we insert resolutions of the identity in the $z$-basis around each exponential of the diagonal Hamiltonian, and resolutions of the identity in the $x$-basis around each off-diagonal to get
\begin{align}
    e^{-\beta(\hat{H}_d + \hat{H}_o)} \approx\!\!\!\!\!\!\! \sum_{z_0,\cdots z_{T-1}, x_0,\cdots x_{T-1}}\!\!\!\!\!\!\! \ket{z_0}
    &\bra{z_0}e^{-\beta \hat{H}_d/T}\ket{z_0}\langle z_0\vert x_0\rangle \bra{x_0} e^{-\beta \hat{H}_o/T}\ket{x_0} \langle{x_0}\vert z_1 \rangle\cdots 
    \\\nonumber
    &\times\langle{z_{T-1}}\vert x_{T-1}\rangle \bra{x_{T-1}} e^{-\beta \hat{H}_o/T} \ket{x_{T-1}}\bra{x_{T-1}}.
\end{align}
Taking the trace, we get that 
\begin{align*}
    \mathcal{Z} &= \mathrm{Tr}\{e^{-\beta \hat{H}}\} \\
    &\approx\!\!\!\!\!\!\! \sum_{z_0,\cdots z_{T-1}, x_0,\cdots x_{T-1}}\!\!\!\!\!\!\! \bra{z_0}e^{-\beta \hat{H}_d/T}\ket{z_0}\langle z_0\vert x_0\rangle \bra{x_0} e^{-\beta \hat{H}_o/T}\ket{x_0} \langle{x_0}\vert z_1 \rangle\cdots \langle{z_{T-1}}\vert x_{T-1}\rangle \bra{x_{T-1}} e^{-\beta \hat{H}_o/T} \ket{x_{T-1}}\langle {x_{T-1}} \vert z_0 \rangle \\
    &=\sum_{z_0,\cdots z_{T-1}, x_0,\cdots x_{T-1}} \prod_{j=0}^{T-1} \bra{z_j}e^{-\beta \hat{H}_d/T}\ket{z_j}\langle z_j\vert x_j\rangle \bra{x_j} e^{-\beta \hat{H}_o/T}\ket{x_{j}} \langle{x_j}\vert z_{j+1} \rangle,
\end{align*}
where $z_T\equiv z_0$ and $x_0\equiv x_T.$ At low spins, it is often not too difficult to explicitly calculate the sum over the $x_j$ variables. However, this quickly becomes unfeasible. As such, we sampled over values of both $x_j$ and $z_j.$

To carry out the Metropolis algorithm, we sweep through the $x_j$ and $z_j$ variables in order, proposing and testing a randomly chosen new value for each dit before moving on to the next.  Each proposal involves a simultaneous proposed change for $z_j$ and the corresponding $x_j$.  In the main text when we report the number of thermalization steps and samples for the Monte Carlo algorithm, each ``step'' of the path integral QMC is a full sweep of proposed changes to every $x_j$ and $z_j$ variable.

\bibliography{main}